\documentclass[onecolumn]{aa}  
\usepackage{graphicx}
\usepackage{txfonts}
\usepackage{comment}
\usepackage{amsmath}
\usepackage{graphicx}
\usepackage{soul}
\usepackage{xcolor}
\usepackage{comment} 
\usepackage{float}
\usepackage{natbib}

\newcommand{\dPgdx}{\frac{\partial\mathcal{P}_{\rm g}}{\partial x}}
\newcommand{\dPgdy}{\frac{\partial\mathcal{P}_{\rm g}}{\partial y}}
\newcommand{\ddPgddx}{\frac{\partial^2\mathcal{P}_{\rm g}}{\partial x^2}}
\newcommand{\ddPgdxdy}{\frac{\partial^2\mathcal{P}_{\rm g}}{\partial x\partial y}}
\newcommand{\ddPgddy}{\frac{\partial^2\mathcal{P}_{\rm g}}{\partial y^2}}
\newcommand{\dddPgdddx}{\frac{\partial^3\mathcal{P}_{\rm g}}{\partial x^3}}
\newcommand{\dddPgddxdy}{\frac{\partial^3\mathcal{P}_{\rm g}}{\partial x^2\partial y}}
\newcommand{\dddPgdxddy}{\frac{\partial^3\mathcal{P}_{\rm g}}{\partial x\partial y^2}}
\newcommand{\dddPgdddy}{\frac{\partial^3\mathcal{P}_{\rm g}}{\partial y^3}}
\newcommand{\ddddPgddddx}{\frac{\partial^4\mathcal{P}_{\rm g}}{\partial x^4}}
\newcommand{\ddddPgdddxdy}{\frac{\partial^4\mathcal{P}_{\rm g}}{\partial x^3\partial y}}
\newcommand{\ddddPgddxddy}{\frac{\partial^4\mathcal{P}_{\rm g}}{\partial x^2\partial y^2}}
\newcommand{\ddddPgdxdddy}{\frac{\partial^4\mathcal{P}_{\rm g}}{\partial x\partial y^3}}
\newcommand{\ddddPgddddy}{\frac{\partial^4\mathcal{P}_{\rm g}}{\partial y^4}}
\newcommand{\effect}{Polarization Leakage}
\begin{document} 
   \title{Polarisation Leakage due to Errors in Track Reconstruction in Gas Pixel Detectors}
   \subtitle{}
   \titlerunning{Polarization leakage in GPDs}
   \author{N. Bucciantini
          \inst{1,2,3}
          \and
          N. Di Lalla\inst{4}
          \and
          R.W.R. Romani\inst{4}
          \and
          S. Silvestri\inst{5, 6}
          \and
          M. Negro\inst{7, 8, 9}
          \and
          L. Baldini\inst{5, 6}
          \and
          A.~F. Tennant\inst{10}
          \and
          A. Manfreda\inst{5}
          }

   \institute{INAF -- Osservatorio Astrofisico di Arcetri, Largo Enrico Fermi 5, 50125 Firenze, Italy\\
        \email{niccolo.bucciantini@inaf.it}
        \and
        Dipartimento di Fisica e Astronomia, Universit\`a degli Studi di Firenze, via G. Sansone 1, Sesto F.no (FI), Italy
        \and
        INFN -- Sezione di Firenze, via G. Sansone 1, Sesto F.no (FI), Italy
        \and
        Department of Physics and Kavli Institute for Particle Astrophysics and Cosmology, Stanford University, Stanford, California 94305, USA.
        \and
        INFN -- Sezione di Pisa, Largo Bruno Pontecorvo 3, Pisa (PI), Italy
        \and
        Dipartimento di Fisica, Universit\`a di Pisa, Largo Bruno Pontecorvo 3, Pisa (PI), Italy
        \and
        University of Maryland, Baltimore County, Baltimore, MD 21250, USA
        \and
        NASA Goddard Space Flight Center, Greenbelt, MD 20771, USA
        \and
        Center for Research and Exploration in Space Science and Technology, NASA/GSFC, Greenbelt, MD 20771, USA
        \and
        NASA Marshall Space Flight Center, Huntsville, AL 35812, USA
        }

   \date{Received .......; accepted ..........}

 
  \abstract
   {X-ray polarimetry, based on Gas Pixel Detectors (GPDs), have reached a high level of maturity with the Imaging X-ray Polarimeter Explorer (IXPE) leading to the first ever spatially resolved polarimetric measures. However, being this a new technique, a few unexpected effect have emerged during in flight operations. In particular it was almost immediately found that on-board unpolarized calibration sources were showing radially polarized halos. The origin of this features was recognized in a correlation between the error in reconstructing the   absorption point of the X-ray photon and the direction of its electric field vector. Here we present and discuss in detail this effect, showing that it is possible to provide a simple and robust mathematical formalism to handle it. We further show its role and relevance for the recent IXPE measures, and for the use of GPD-based techniques in general, and illustrate how to model it in the study of extended sources. }

   \keywords{ Instrumentation: detectors -- Instrumentation: polarimeters -- Techniques: polarimetric -- X-rays: general -- Polarization   }

   \maketitle
%

\section{Introduction}
Since the launch of the first X-ray satellites in the '60s \citep{Kalemci18a}, X-ray astronomy has provided us a unique window to investigate high-energy astrophysical environments, such as Pulsar \citep{Walter_Ferrigno17a}, Magnetars \citep{Mereghetti_Pons+15a}, Pulsar Wind Nebulae  \citep{Kargaltsev_Pavlov08a}, Supernova Remnants  \citep{Vink12a}, Black Hole  \citep{Singh13a,Bachetti16}, Clusters \citep{Sarazin86a}, and Accretion Disks  \citep{Inoue22}. X-ray emission in many of those systems is typically a mixture of thermal and non-thermal components, with the latter due essentially to synchrotron radiation from accelerated particles (pairs) in the strong local magnetic field. Pulsar Wind Nebulae, and Supernova Remnants are resolved sources showing such emission. One of the key feature of synchrotron emission is its high level of polarization \citep{Bucciantini18a}. Polarization might also arise in thermal environments due to plasma propagation effect and for strongly magnetized systems, like Magnetars, due to QED \citep{Turolla_Zane+15a}.\\
\\
Unfortunately while X-ray imaging and spectral instruments have been operating for a long time, the same cannot be said for polarimetry. Until very recently the only X-ray source where X-ray polarimetric measurements have been firmly established was the Crab Nebula \citep{Weisskopf_Silver+78a}.  With the recent launch of the Imaging X-ray Polarimeter Explorer (IXPE) \citep{IXPEcalibration}, this gap is finally filled, offering us for the fist time the opportunity to obtain space resolved polarimetric observations of several extended Galactic sources, and even some extra-Galactic objects.\\
\\
The polarization sensitive Gas Pixel Detectors (GPDs) on board IXPE  [and previously used also by PolarLight, \citet{Feng+19a}]exploit the properties of the photoelectric effect \citep{articleCosta}: the photon, focused by a mirror system, and collimated onto the sensitive detector area, converts in a low-Z gas of pure dimethyl ether, emitting a photoelectron (PE) that produces an ionization track. This process is quite sensitive to photon polarization given that the initial direction of emitted PE is preferentially in the direction of the photon electric field (polarization vector, EVPA), and has large cross sections in the low energy range [2-10]~keV, where many of the most interesting astrophysical X-ray sources shine.\\
\\
Key to derive the polarization of the X-ray photon, is to reconstruct the initial direction of the PE track. This however is prone to lead to spurious effects, whenever the photo-electron tracks are not perfectly reconstructed. In particular, failure to reconstruct the exact conversion point (the absorption site) of the incoming photon, can create false polarization patterns, that are particularly prominent for sources characterized by sharp edges. This first emerged clearly, soon after the launch of the IXPE satellite, as a radial polarization pattern seen around bright compact unpolarized (on board) calibration sources, especially in the low intensity wings of the sources themselves,  and later found also in unpolarized sky point sources.\\
\\
Here we discuss this phenomenon and present a mathematical formalization, showing that in analogy with what is done in radio astronomy, where the \effect, mainly due to non-ideality of the antennas, is well known \citep{Hales17a}, is possible to model it in terms of Stokes-dependent Point Spread Functions (PSFs), which can be related to the Mueller Matrix elements \citep{Tinbergen96a}, that provide the full imaging response of an instrument to polarized radiation.\\
\\
In Sect.~\ref{sec:gpd} we review the working of a GPD, and how tracks are reconstructed. In Sect.~\ref{sec:heuristic} we present an heuristic explanation of the origin of \effect, and in Sect.~\ref{sec:formalism} we introduce a mathematical formalization of the effects, and verify the limits of our formalism in predicting these effects, which we later compared with either true IXPE data or Monte Carlo simulations of the GPD in Sect.~\ref{sec:comp}. In Sect.~\ref{sec:mueller} we illustrate how the analysis can be framed in terms of Mueller Matrix elements, and illustrate some applications. We finally present our conclusions in Sect.~\ref{sec:conclusion}.


\section{Tracks Reconstruction in Gas Pixel Detectors}
\label{sec:gpd}
\begin{figure}[ht]
    \centering
    \includegraphics[width=7cm, height=7cm]{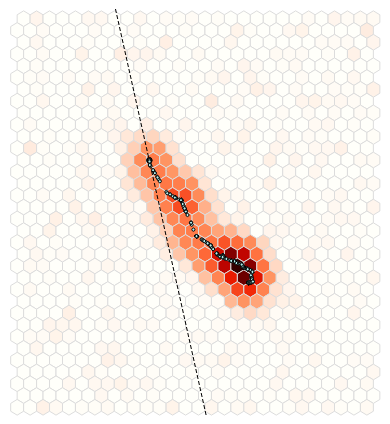}
    \includegraphics[width=7cm, height=7cm]{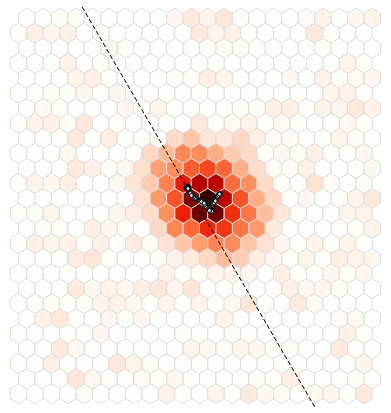}
    \caption{Example of a photoelectron track generated by a 7 keV photon (left panel) and a 3.5 keV photon (right panel). The dashed lines and the black points represent respectively the simulated PE emission direction and X-ray impact point. The simulated photoelectron path is reported as well. Ionization charges created along the photoelectron path are reported as small cyan dots. The color scale, going from white to dark red, represents the amount of charge collected in each pixel 
    }
    \label{fig:tracks}
\end{figure}

\begin{figure}[ht]
    \centering
    \includegraphics[width=8cm, height=7cm, clip, bb=150 150 1100 1100]{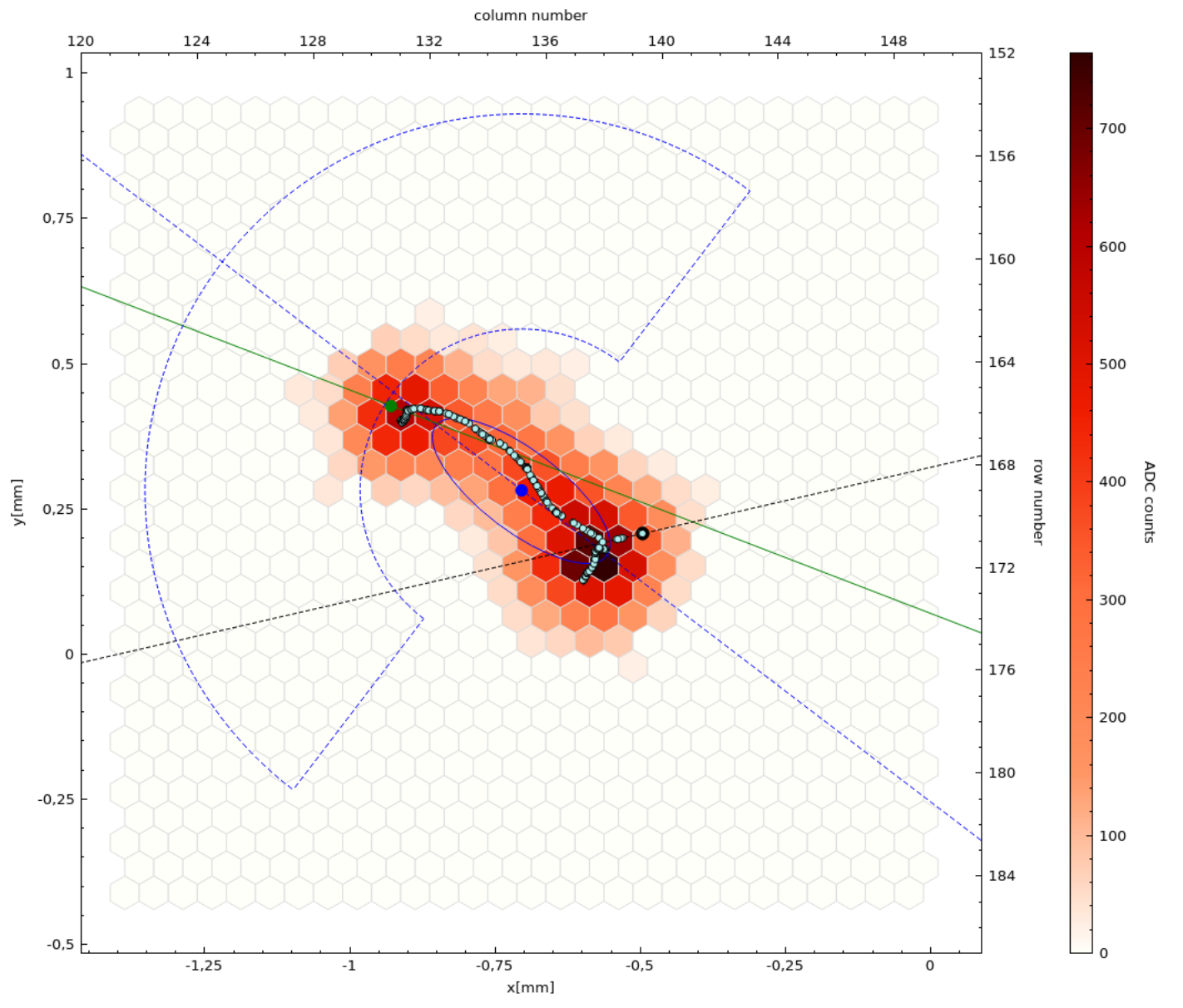}
\caption{Example of a wrongly reconstructed absorption point. The black dot is the real absorption point, the green one is the reconstructed absorption point and the blue one is the barycenter. Note that the reconstructed polarization plane (blues dashed line) differs from the true one (black dashed line). The colors have the same meaning as in Fig.~\ref{fig:tracks}.
    }
    \label{fig:wrong}
\end{figure}

X-ray polarimetry exploiting the high dependence of the photoelectric effect on the polarization of the incident radiation can be substantially more efficient than previously adopted techniques [i.e. Bragg reflection and Thomson scattering -- these both suffer from limitations \citep{Soffitta_PolMeasReview} of energy band width, lack imaging capabilities or offer low effective area]. An example of detector design based on this property has been proposed and developed in early 2000 \citep{articleCosta}, and reached a mature state with the selection of IXPE \citep{IXPEcalibration}. This instrument combines good imaging capabilities and unprecedented polarization sensitivities exploiting a GPD \citep{BellazziniGPD} design.\\
\\
An X-ray photon is absorbed in the gas gap of the GPD, and a photoelectron  is ejected in the direction ($\theta, \phi$), which preferably lies on the oscillation plane of the electric field of the incoming X-ray (namely the polarization direction). $\theta$ is the angle between the incident X-ray direction and the PE emission direction, while $\phi$ is the azimuthal PE emission direction \citep{Sabbatucci_Salvat16a}. The photoelectron interacts with the gas atoms through ionizing collisions, loosing energy and  changing direction due to multiple scattering, with the energy losses that increase at each collision\footnote{The energy loss $\frac{\partial E}{\partial x}$ is inversely proportional to the kinetic energy of the electron: $\frac{\partial E}{\partial x} \propto \frac{1}{\beta^2} \propto \frac{1}{E_{{\rm }kin}}$, where $\beta$ is the velocity of the electron and $E_{{\rm }kin}$ is its kinetic energy, so that the most of the energy is lost at the end of the track (Bragg peak).}. Such interactions generate a pattern of ions-e$^-$ that mark the path followed by the PE before it loses all its energy and is reabsorbed in the gas. Such a pattern, typically a few hundred micron across, is called a \textit{track}. The ions-e$^-$ charges are amplified by a Gas Electron Multiplier (GEM) and collected on a plane of hexagonal pixels in an honeycomb configuration. A track, therefore, is a pixelated image containing useful information about the PE, and, as a consequence, about the X-ray that generated that same PE. Two examples of a PE track image for two different X-ray energies are reported in Fig.~\ref{fig:tracks}.\\
\\
The event reconstruction consists in the estimation of the properties of interest of the incoming X-ray photon from the PE track image. In particular, the total collected charge is proportional to the energy of the X-ray, the starting point of the track gives the impact point of the photon, and the azimuthal angle $\phi$ of the PE initial emission direction (before it gets deviated by multiple interactions in the gas) carries the imprint of the X-ray polarization direction. As shown if Fig.~\ref{fig:tracks}, the track morphology depends on the energy of the absorbed X-ray; in general it results more challenging to reconstruct the initial PE direction for low energy X-rays due to the almost circular shape of the track. \\
\\
The event reconstruction is typically done analytically, processing the track image to estimate both the impact point of the X-ray and the PE initial direction. A detailed description of the analytic algorithm used in IXPE event reconstruction can be found in \citet{AnalyticRecon}. In short, after a zero suppression and clustering stage to identify and isolate the main track, the reconstruction algorithm proceeds with a moment analysis of the two-dimensional charge distribution. The whole procedure is iterative and, after the initial determination of the barycenter and the principal axis of the track, the moment analysis is run again focusing on the head of the track, supposedly containing the impact point. Most of the energy is indeed released at the end of the photoelectron track (Bragg peak) and thus the projection of the charge distribution along the principal axis allows us to distinguish the head and the tail of track. As an alternative to the analytic algorithm, machine learning techniques can be used. In particular, convolutional Neural Networks (CNNs) have been explored and tested on simulated data, showing promising results \citep{TakaoCNNclassification, CNNregression, StanfordCNN, ROMANI2021}.\\
\\
In Fig.~\ref{fig:wrong} we illustrate a typical example of mistakes that can take place in the reconstruction of the property of a track. In this case the absorption point of the incoming X-ray photon, has been reconstructed at the end of the track, close to the Bragg peak, and not at its beginning (due to the intrinsically 3D nature of the PE emission process, some of the ambiguity in the inferred absorption point could be due
to initial photo-electron coming out in a more vertical direction). This often occurs when an Auger electron deposits substantial energy near the conversion point. In fact, the error in the conversion point is generally largest along the major axis of the track, which correlates with the EVPA. As we discuss in the next section, it is this kind of error that leads to the \effect, that was first observed in IXPE calibration data. Note also that the reconstructed polarization plane, does not correspond to the true direction. This is due in part to the fact that the PE is aligned with the photon electric field only on average, and in part due to errors in properly reconstructing the track shape. The fact that the reconstructed polarization plane matches the true polarization plane only on average, implies that the efficiency of the polarized response of a GPD based detector, the so called \textit{modulation factor} can be substantially smaller than unity. 


\section{Heuristic explanation}
\label{sec:heuristic}

\begin{figure}[H]
    \centering
    \includegraphics[width=17cm, clip, bb=60 150 750 490]{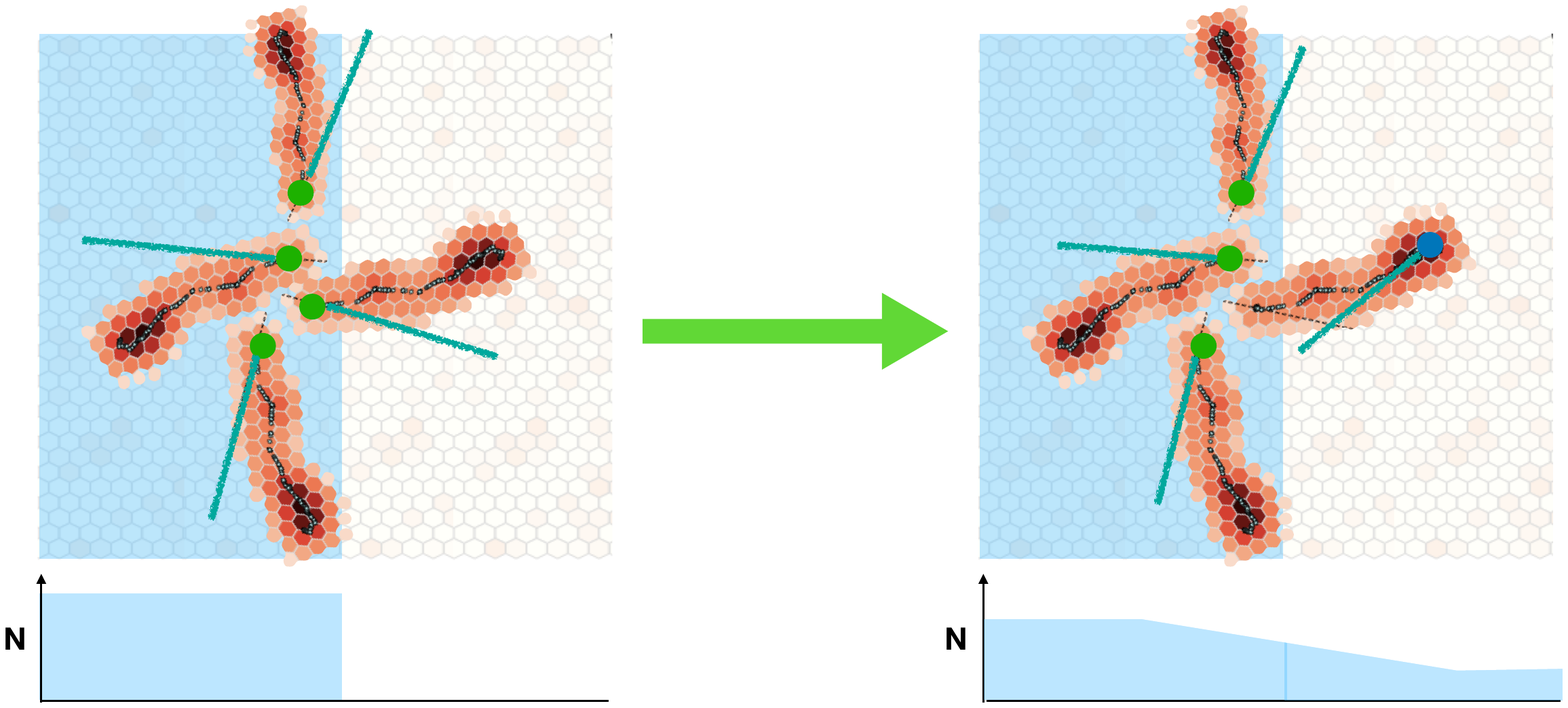}
    \caption{Schematic picture of \effect~ for an unpolarised source with a sharp edge. The light blue region represents the portion of the GPD corresponding to the source. Four events, close to the edge of the source image, are represented with their respective PE tracks. The green dots represent the true absorption points and the blue segments are the reconstructed polarization planes of the original photons. On the left all absorption points are correctly reconstructed in their true positions: all events are inside the edge, and none outside (bottom panel represents the event count); there are as many PEs with tracks tangential as orthogonal to the edge so no net polarization is assigned to the source.  On the right the absorption point of the track extending outside the source edge has been wrongly reconstructed at the end of the track (blue dot): there are less events reconstructed inside the edge, and more outside (bottom panel represents the event count); there are  more PEs reconstructed with tracks tangential than orthogonal to the edge inside the source region, while there are PE events with tracks orthogonal to the edge outside. This leads to a blurring of the edge and to a polarization pattern tangential inside the edge and orthogonal outside. }
    \label{fig:heuristic}
\end{figure}

In order to briefly explain how reconstruction errors in the absorption point lead to \effect~ let us consider a source characterized by a sharp edge (see Fig.~\ref{fig:heuristic}). 
Recall that all the GPD \textit{measures} is the charge collected along the full track: both the impact point and the polarization direction are reconstructed. In a source with a sharp edge, as in figure \ref{fig:heuristic}, there will be tracks that extend beyond the edge and for these there will be a chance of reconstructing the impact point outside the true source boundary. This not only will assign a photon to a region where there should be none, thus blurring the edge, but will systematically assign to those photons a polarization direction normal to the boundary. This will cause an excess of orthogonally polarized photon outside the edge, and consequently an excess of tangentially polarized photons inside it. Note that, {\it on average} this effect does not change the inferred polarization plane of the photon, given that the impact point is typically misplaced along the photo-electron track (the principal axis of the track, which correlated with the incoming photon polarization plane).\\
\\
Even if a source does not have such a sharp boundary but a simple gradient of the flux, changing significantly on the scale length of the typical extent of the tracks, there will be a polarization gradient. The result is a polarized `halo' around any source, even if there is no intrinsic polarization. For point sources these radial polarization gradients average out to no net polarization when integrated around the source at all azimuths. However for extended sources, or faint sources near bright sources, these induced polarization structures can corrupt the true polarization signal.\\
\\
Note that this effect is intrinsic of any polarization measure based on track reconstruction, as long as there is a correlation between the offset of the absorption point from its true position and the inferred polarization direction. This is inevitable given that any position uncertainty will be largest along the long axis of the track. Since this axis is, especially for well-measured higher energy photons, along the estimated EVPA, there is naturally a strong correlation between the position offset direction and the inferred polarization.\\  
\\
By looking at Fig.~\ref{fig:heuristic} it is clear that for simple sources, characterized by edges, it might be possible to discriminate between a true intrinsic polarization, and a polarized feature due to \effect. For a truly polarized signal the chance of a PE track having an azimuthal angle $\phi$ is the same as having an azimuthal angle $\phi \pm \pi$ (due to the very nature of the PE emission), and the distribution function of the azimuthal angles of the tracks will be dominated by a $m=2$ mode. In a polarised halo due to \effect~ all tracks will tend to point toward the edge (anti-aligned to intensity gradients), and the distribution function of the azimuthal angles of the tracks will be dominated by a $m=1$ mode. However, for real sources, characterized by complex intensity patterns, such simple characterization might not apply. Unfortunately this information is completely lost when transforming from the PE azimuthal angle $\phi$ to the photon Stokes parameters given that $Q(\phi) = Q(\phi \pm \pi)$ (and the same for $U$). This is an issue in IXPE level 2 files given that they only provide the event Stokes parameters (IXPE processing corrects instrument effects
in Q,U space; it is not easy to correct the inferred electron track
provided in the level 1 files) and not the original PE angle. \\
\\


\section{Formalization}
\label{sec:formalism}
In principle, the offset of the reconstructed absorption point from the true absorption point will have components both along and transverse to the reconstructed polarization plane. In order to simplify the analysis we will assume for simplicity that the offset is only along the reconstructed polarization plane (purely one dimensional). A more general treatment can be found in Appendix~\ref{sec:general}, where a generic correlation is assumed. Interestingly, as the comparison with the more general results shows, our simplified model captures very well the properties of the \effect, to the point that it provides reliable estimates of true data.

\subsection{One-dimensional displacement}
\begin{figure}[H]
    \centering
    \includegraphics[width=8cm]{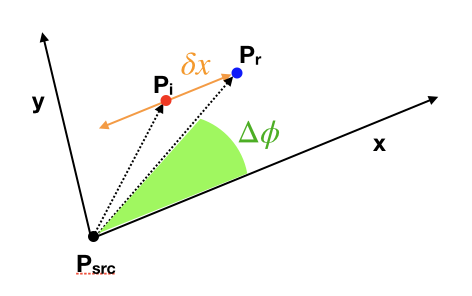}
    \caption{Schematic representation of the geometry involved in the absorption point displacement due to reconstruction errors. $\mathbf{P}_{\rm src}$ is the location of the point source whence the photon come from. $\mathbf{P}_{\rm i}$ is the true impact (absorption) point of the photon, displaced due to the mirror PSF, while the orange double-headed arrow represents the photo-electron direction $\mathbf{n}$. $\mathbf{P}_{\rm r}$ is the reconstructed impact point displaced by an amount $\delta x$ along the photo-electron direction plane. $x-y$ are coordinate axes chosen such that $x$ is parallel to the photo-electron direction plane, while $\Delta \phi$ is the polar angle of the reconstructed impact point in the same reference frame (describes the azimuthal distribution of photons due to the mirror PSF.}
    \label{fig:fig1}
\end{figure}

Let us assume that a photon, from a point source, located at $\mathbf{P}_{\rm src}$ (see Fig.~\ref{fig:fig1}) interacts with the detector at a point $\mathbf{P}_{\rm i}$ on the detector plane, and produces a photo-electron track in a direction $\mathbf{n}$. The photo-electron direction represents the measured polarization plane of the photon. Now consider a reference frame centered on the position of the point source and with the $x$-axis aligned with the direction $\mathbf{n}$. In this reference frame the interaction point (the absorption or impact point) will be located at $\mathbf{P}_{\rm i}=(x_{\rm i},y_{\rm i})$. The chance of finding a photon at a location $(x,y)$ is the point-spread-function (PSF) of the instrument $\mathbf{P}_{\rm g}(x,y)$, where with the subscript ``${\rm g}$'' we imply that this is the geometrical-PSF, i.e. the PSF of a light ray.  It might happen that, due to uncertainties in the reconstruction of the track shape, the interaction point of the photon instead of being reconstructed in its true position $\mathbf{P}_{\rm i}$, is reconstructed at a position displaced along  $\mathbf{n}$. In our reference frame the reconstructed impact point will be located at $\mathbf{P}_{\rm r}=(x_{\rm i}+\delta x,y_{\rm i})$.  Let it be $\mathcal{P}_{\rm f}(\delta x)$ the probability distribution function of the absorption point displacement due to reconstruction errors, which can be safely assumed to be independent on the impact position. It is also safe to assume that $\mathcal{P}_{\rm f}(\delta x)=\mathcal{P}_{\rm f}(-\delta x)$, i.e.  there is no preference in the sign of the displacement, due to the very nature of the photo-electron emission process.\\

Then one can define a total PSF arising from the combined geometrical and polarization-shifting offsets. The total chance of reconstructing a photon impact point at a location $(x,y)$ will be:
\begin{align}
    \mathcal{P}_{\rm t}(x,y) &=\int_{-\infty}^{\infty} \!\!\!\mathcal{P}_{\rm g}(x+\delta x,y)\mathcal{P}_{\rm f}(-\delta x) d \delta x \label{eq:1}
     =\mathcal{P}_{\rm g}(x,y) +\sum_{n=1}^{\infty}\frac{1}{(2n)!}\left(\frac{\partial^{2n} \mathcal{P}_{\rm g}}{\partial x^{2n}} \right)\int_{-\infty}^{\infty}\!\!\! (\delta x)^{2n}\mathcal{P}_{\rm f}(\delta x) d \delta x
     =\mathcal{P}_{\rm g}(x,y) +\sum_{n=1}^{\infty}\frac{1}{(2n)!}\left(\frac{\partial^{2n} \mathcal{P}_{\rm g}}{\partial x^{2n}} \right)m_{\rm f}^{2n}
\end{align}
where $m_{\rm f}^{2n}$ is the central moment of order $2n$ of $\mathcal{P}_{\rm f}(\delta x)$ ($m_{\rm f}^{2}$ is by definition the central variance of the distribution $\sigma_{\rm f}^2$). For examples, in the case in which  $\mathcal{P}_{\rm f}(\delta x)$ is a Gaussian with variance $\sigma_{\rm f}^2$ then one has:
\begin{align}
\mathcal{P}_{\rm t}(x,y)=\mathcal{P}_{\rm g}(x,y) +\sum_{n=1}^{\infty}\frac{(2n-1)!!}{(2n)!}\left(\frac{\partial^{2n} \mathcal{P}_{\rm g}}{\partial x^{2n}} \right)\sigma_{\rm f}^{2n}
\end{align}
If one further assumes that also the geometrical PSF is Gaussian with variance $\sigma_{\rm g}$ then one finds:
\begin{align}
\mathcal{P}_{\rm t}(x,y)=\frac{1}{2\pi\sigma_{\rm g}^2}e^{-(x^2+y^2)/2\sigma_{\rm g}^2}\times \sum_{n=0}^{\infty}\frac{(2n-1)!!}{(2n)!}H_{2n}(x/\sigma_{\rm g})\sigma_{\rm f}^{2n}
\end{align}
where $H_{2n}(x)$ is the Hermite polynomial of order $2n$.\\
\\
\subsection{Linearized Theory}
\label{sec:linear}
In the limit $\sigma_{\rm f} \ll \mathcal{P}_{\rm g}/|\nabla \mathcal{P}_{\rm g}|$, one can approximate as:
\begin{align}
\mathcal{P}_{\rm t}(x,y)=\mathcal{P}_{\rm g}(x,y) +\frac{1}{2}\left(\frac{\partial^{2} \mathcal{P}_{\rm g}}{\partial x^{2}} \right)\sigma_{\rm f}^{2} + \mathcal{O}(\sigma_{\rm f}^{4})\label{eq:eq6}
\end{align}
If the geometrical-PSF is radially-symmetric $\mathcal{P}_{\rm g}(x,y) = \mathcal{P}_{\rm g}(r)$ with $r=\sqrt{(x^2+y^2)}$, then one has:
\begin{align}
    \frac{\partial^{2} \mathcal{P}_{\rm g}}{\partial x^{2}} = \frac{\partial^{2} \mathcal{P}_{\rm g}}{\partial r^{2}}\left(\frac{x}{r} \right)^2+\frac{1}{r}\frac{\partial \mathcal{P}_{\rm g}}{\partial r}\left(\frac{y}{r} \right)^2 \label{eq:eq7}
\end{align}
introducing the relative angle $\Delta \phi$ between the vector-position of the measured impact point $\mathbf{P}_{\rm r}$ and the photo-electron direction $\mathbf{n}$ (which describes the azimuthal spread of photons due to the mirror PSF, and as such is totally uncorrelated with the displacement due to reconstruction error), one has:
\begin{align}
    \frac{\partial^{2} \mathcal{P}_{\rm g}}{\partial x^{2}} &= \frac{\partial^{2} \mathcal{P}_{\rm g}}{\partial r^{2}}\cos^2{(\Delta \phi)}+\frac{1}{r}\frac{\partial \mathcal{P}_{\rm g}}{\partial r}\sin^2{(\Delta \phi)} \nonumber
     = \frac{\partial^{2} \mathcal{P}_{\rm g}}{\partial r^{2}}\frac{1+\cos{(2\Delta \phi)}}{2}+\frac{1}{r}\frac{\partial \mathcal{P}_{\rm g}}{\partial r}\frac{1-\cos{(2\Delta \phi)}}{2} \nonumber\\
     &= \frac{1}{2}\left[\frac{\partial^{2} \mathcal{P}_{\rm g}}{\partial r^{2}}+\frac{1}{r}\frac{\partial \mathcal{P}_{\rm g}}{\partial r}\right]+\frac{1}{2}\left[\frac{\partial^{2} \mathcal{P}_{\rm g}}{\partial r^{2}}-\frac{1}{r}\frac{\partial \mathcal{P}_{\rm g}}{\partial r}\right]\cos{(2\Delta \phi)}
\end{align}
hence
\begin{align}
\mathcal{P}_{\rm t}(x,y)=\mathcal{P}_{\rm g}(r)+\frac{\sigma_{\rm f}^2}{4}\left[\frac{\partial^{2} \mathcal{P}_{\rm g}}{\partial r^{2}}+\frac{1}{r}\frac{\partial \mathcal{P}_{\rm g}}{\partial r}\right]+\frac{\sigma_{\rm f}^2}{4}\left[\frac{\partial^{2} \mathcal{P}_{\rm g}}{\partial r^{2}}-\frac{1}{r}\frac{\partial \mathcal{P}_{\rm g}}{\partial r}\right]\cos{(2\Delta \phi)} + \mathcal{O}(\sigma_{\rm f}^{4})
\end{align}
where $x =r\cos{(\Delta \phi)}$ and $y =r\sin{(\Delta \phi)}$. The effect of misplacing the absorption point is twofold: on one hand it modifies the radially symmetric part of the PSF, on the other it adds a modulation along the photo-electron polarization plane. The total PSF is now a function also of the polarization properties of the photon under consideration.
\\
\\
\subsubsection{Unpolarized sources}

Let us now consider an unpolarized point sources. The chance that a photon has a measured impact point at location $(x,y)$ is:
\begin{align}
\mathcal{P}_{\rm I}(x,y)= \int_0^{2\pi} \mathcal{P}_{\rm t}(x,y)\mathcal{F}({\Delta \phi})d\Delta\phi
\end{align}
where $\mathcal{F}({\Delta \phi})$ is the  photo-electron direction distribution function of the photons coming from the source, i.e. the chance that a photon from the point sources has a polarization plane at an angle $\Delta \phi$ (in the range $[0,2\pi]$) with respect to the vector position defined by $(x,y)$. For an unpolarized source $\mathcal{F}({\Delta \phi}) = 1/2\pi$, and one has:
\begin{align}
\mathcal{P}_{\rm I}(x,y)=\mathcal{P}_{\rm I}(r) =\mathcal{P}_{\rm g}(r)+\frac{\sigma_{\rm f}^2}{4}\left[\frac{\partial^{2} \mathcal{P}_{\rm g}}{\partial r^{2}}+\frac{1}{r}\frac{\partial \mathcal{P}_{\rm g}}{\partial r}\right] +\mathcal{O}(\sigma_{\rm f}^4)
\end{align}
which represents the PSF of the Intensity field of an unpolarized point source. An unpolarized point source with intensity $I_{\rm src}$ will give, to first order in $\sigma_{\rm f}^2$, an image with intensity:
\begin{align}
I(x,y) = I_{\rm src}\left\{\mathcal{P}_{\rm g}(r)+\frac{\sigma_{\rm f}^2}{4}\left[\frac{\partial^{2} \mathcal{P}_{\rm g}}{\partial r^{2}}+\frac{1}{r}\frac{\partial \mathcal{P}_{\rm g}}{\partial r}\right]\right\}
\end{align}
showing that this effect leads to an additional blurring of the image.
\\
One can also look at the Stokes parameters $Q$ and $U$. This however requires to choose an absolute reference frame ($Q$ and $U$ are always defined with respect to an absolute frame). We set $x =r\cos{(\phi_{\rm l})}$ and $y = r\sin{(\phi_{\rm l})}$, such that $\Delta \phi = \phi_{\rm l} -\phi$ where $\phi$ is the photo-electron angle measured in the same absolute frame. The value of the Stokes parameters at a point $(x,y)$ can then be computed according to:
\begin{align}
  \mathcal{P}_{\rm Q}(x,y) = 2\int_0^{2\pi} \cos{(2\phi)}\mathcal{P}_{\rm t}(x,y)\mathcal{F}d\phi,\quad\quad\quad 
  \mathcal{P}_{\rm U}(x,y) = 2\int_0^{2\pi} \sin{(2\phi)}\mathcal{P}_{\rm t}(x,y)\mathcal{F}d\phi
\end{align}
where $\mathcal{P}_{\rm Q}$ and $\mathcal{P}_{\rm U}$ can be thought of as the PSF for the Stokes parameters, thus generalizing the definition of PSF that typically refers only to the Intensity.\\
\\
One finds to $\mathcal{O}(\sigma_{\rm f}^4)$ accuracy:
\begin{align}
\mathcal{P}_{\rm Q}(x,y)&= 2\int_0^{2\pi} \!\!\!\!\frac{\sigma_{\rm f}^2}{4}\left[\frac{\partial^{2} \mathcal{P}_{\rm g}}{\partial r^{2}}-\frac{1}{r}\frac{\partial \mathcal{P}_{\rm g}}{\partial r}\right]\cos{(2\phi)}\cos{(2(\phi_{\rm l} -\phi))}\frac{1}{2\pi}d\phi =\frac{\sigma_{\rm f}^2}{4}\left[\frac{\partial^{2} \mathcal{P}_{\rm g}}{\partial r^{2}}-\frac{1}{r}\frac{\partial \mathcal{P}_{\rm g}}{\partial r}\right]\cos{(2\phi_{\rm l})} \nonumber\\
&
=\frac{\sigma_{\rm f}^2}{4}\left[\frac{\partial^{2} \mathcal{P}_{\rm g}}{\partial r^{2}}-\frac{1}{r}\frac{\partial \mathcal{P}_{\rm g}}{\partial r}\right]\frac{x^2-y^2}{r^2}
\end{align}
and
\begin{align}
\mathcal{P}_{\rm U}(x,y)&= 2\int_0^{2\pi} \!\!\!\!\frac{\sigma_{\rm f}^2}{4}\left[\frac{\partial^{2} \mathcal{P}_{\rm g}}{\partial r^{2}}-\frac{1}{r}\frac{\partial \mathcal{P}_{\rm g}}{\partial r}\right]\sin{(2\phi)}\cos{(2(\phi_{\rm l} -\phi))}\frac{1}{2\pi}d\phi =\frac{\sigma_{\rm f}^2}{4}\left[\frac{\partial^{2} \mathcal{P}_{\rm g}}{\partial r^{2}}-\frac{1}{r}\frac{\partial \mathcal{P}_{\rm g}}{\partial r}\right]\sin{(2\phi_{\rm l})}\nonumber\\
&
=\frac{\sigma_{\rm f}^2}{4}\left[\frac{\partial^{2} \mathcal{P}_{\rm g}}{\partial r^{2}}-\frac{1}{r}\frac{\partial \mathcal{P}_{\rm g}}{\partial r}\right]\frac{2xy}{r^2}
\end{align}
which is exactly what is expected for a pattern of radial and/or tangential polarization. In particular one expects radial vs tangential polarization depending of the sign of the term in square brackets.\\
\\
\\
\\
\subsubsection{Polarized sources}

What we investigated in the previous section was the effect the  displacement of the absorption point due to reconstruction errors, on the Intensity distribution and the Stokes $U$ and $Q$ distributions, for an unpolarized point source. In a similar fashion one can repeat the computation for a polarized point source.\\
\\
For polarized sources it is more convenient to use always an absolute reference frame. As before, we take an absolute reference frame where $x =r\cos{(\phi_{\rm l})}$ and $y = r\sin{(\phi_{\rm l})}$, and $\Delta \phi = \phi_{\rm l} -\phi$ where $\phi$ is the photo-electron angle measured in the same absolute frame. A polarized source, will have a photo-electron distribution given by:
\begin{align}
\mathcal{F}(\phi) = [1 + a_{\rm src}\cos{(2(\phi-\phi_o))}]/2\pi
\end{align}
The Stokes parameter of the source are defined as:
\begin{align}
Q_{\rm src}/I_{\rm src}  = 2  \int_0^{2\pi} \!\!\! \mathcal{F}(\phi) \cos{(2\phi)} d\phi \quad\quad U_{\rm src}/I_{\rm src}  = 2  \int_0^{2\pi} \!\!\! \mathcal{F}(\phi) \sin{(2\phi)} d\phi 
\end{align}  
which provides the following relation between the  the amplitude of the modulation $a_{\rm src}$  and the Stokes parameters: 
$a_{\rm src} = \sqrt{Q_{\rm src}^2 + U_{\rm src}^2}/I_{\rm src}$, with $Q_{\rm src}/I_{\rm src} = a_{\rm src}\cos{(2\phi_o)}$, and $U_{\rm src}/I_{\rm src} = a_{\rm src}\sin{(2\phi_o)}$. \\
\\
Then one finds to $\mathcal{O}(\sigma_{\rm f}^4)$ accuracy:
\begin{align}
\mathcal{P}_{\rm I}(x,y)&=\mathcal{P}_{\rm g}(r)+\frac{\sigma_{\rm f}^2}{4}\left[\frac{\partial^{2} \mathcal{P}_{\rm g}}{\partial r^{2}}+\frac{1}{r}\frac{\partial \mathcal{P}_{\rm g}}{\partial r}\right] 
+ \int_0^{2\pi} \!\!\!\!a_{\rm src}\frac{\sigma_{\rm f}^2}{4}\left[\frac{\partial^{2} \mathcal{P}_{\rm g}}{\partial r^{2}}-\frac{1}{r}\frac{\partial \mathcal{P}_{\rm g}}{\partial r}\right]\cos{(2(\phi_{\rm l} -\phi))}\frac{\cos{(2(\phi-\phi_o))}}{2\pi}d\phi\\
&=\mathcal{P}_{\rm g}(r)+\frac{\sigma_{\rm f}^2}{4}\left[\frac{\partial^{2} \mathcal{P}_{\rm g}}{\partial r^{2}}+\frac{1}{r}\frac{\partial \mathcal{P}_{\rm g}}{\partial r}\right]+\frac{a_{\rm src}}{2}\frac{\sigma_{\rm f}^2}{4}\left[\frac{\partial^{2} \mathcal{P}_{\rm g}}{\partial r^{2}}-\frac{1}{r}\frac{\partial \mathcal{P}_{\rm g}}{\partial r}\right]\cos{(2(\phi_{\rm l}-\phi_o))}
\end{align}
or
\begin{align}
\mathcal{P}_{\rm I}(r\cos{(\phi_{\rm l})},r\sin{(\phi_{\rm l})})&=\mathcal{P}_{\rm g}(r)+\frac{\sigma_{\rm f}^2}{4}\left[\frac{\partial^{2} \mathcal{P}_{\rm g}}{\partial r^{2}}+\frac{1}{r}\frac{\partial \mathcal{P}_{\rm g}}{\partial r}\right]+\frac{a_{\rm src}}{2}\frac{\sigma_{\rm f}^2}{4}\left[\frac{\partial^{2} \mathcal{P}_{\rm g}}{\partial r^{2}}-\frac{1}{r}\frac{\partial \mathcal{P}_{\rm g}}{\partial r}\right]\cos{(2(\phi_{\rm l}-\phi_o))}
\end{align}
which in terms of the source Stokes parameters reads:
\begin{align}
\mathcal{P}_{\rm I}(x,y)&=\mathcal{P}_{\rm g}(r)+\frac{\sigma_{\rm f}^2}{4}\left[\frac{\partial^{2} \mathcal{P}_{\rm g}}{\partial r^{2}}+\frac{1}{r}\frac{\partial \mathcal{P}_{\rm g}}{\partial r}\right]+\frac{\sigma_{\rm f}^2}{8}\left[\frac{\partial^{2} \mathcal{P}_{\rm g}}{\partial r^{2}}-\frac{1}{r}\frac{\partial \mathcal{P}_{\rm g}}{\partial r}\right][Q_{\rm src}\cos{(2\phi_{\rm l})}+U_{\rm src}\sin{(2\phi_{\rm l})}]/I_{\rm src}
\end{align}
now $\sin{(2\phi_{\rm l})} = 2xy/r^2$ while $\cos{(2\phi_{\rm l})} = x^2/r^2-y^2/r^2$ which finally allows one to write:
\begin{align}
\mathcal{P}_{\rm I}(x,y)&=\mathcal{P}_{\rm g}(r)+\frac{\sigma_{\rm f}^2}{4}\left[\frac{\partial^{2} \mathcal{P}_{\rm g}}{\partial r^{2}}+\frac{1}{r}\frac{\partial \mathcal{P}_{\rm g}}{\partial r}\right]+\frac{\sigma_{\rm f}^2}{8}\left[\frac{\partial^{2} \mathcal{P}_{\rm g}}{\partial r^{2}}-\frac{1}{r}\frac{\partial \mathcal{P}_{\rm g}}{\partial r}\right]  \left[\frac{Q_{\rm src}}{I_{\rm src}}\frac{x^2-y^2}{r^2}+\frac{U_{\rm src}}{I_{\rm src}}\frac{2xy}{r^2} \right]
\end{align}
which shows that for polarized sources, the PSF is elongated along the direction of the polarization. In general the PSF will have 3 contributions: the leading geometrical symmetric PSF, a symmetric term due to unpolarized radiation, and an elongated term associated with the polarized component.
\\
Looking at the Stokes parameters $Q$ and $U$, and recalling that
\begin{align}
    \int_0^{2\pi}\cos{(2\phi)}\cos{(2(\phi_{\rm l} -\phi))}\cos{(2(\phi-\phi_{\rm o}))}d\phi = 0\quad{\rm and}\quad
    \int_0^{2\pi}\sin{(2\phi)}\cos{(2(\phi_{\rm l} -\phi))}\cos{(2(\phi-\phi_{\rm o}))}d\phi = 0
\end{align}
one can easily show that for a point source with non-zero $Q_{\rm src}$ and $U_{\rm src}$
\begin{align}
\mathcal{P}_{\rm Q}(x,y)
&=\frac{Q_{\rm src}}{I_{\rm src}}\left\{\mathcal{P}_{\rm g}(r)+\frac{\sigma_{\rm f}^2}{4}\left[\frac{\partial^{2} \mathcal{P}_{\rm g}}{\partial r^{2}}+\frac{1}{r}\frac{\partial \mathcal{P}_{\rm g}}{\partial r}\right]\right\}+ \frac{\sigma_{\rm f}^2}{4}\left[\frac{\partial^{2} \mathcal{P}_{\rm g}}{\partial r^{2}}-\frac{1}{r}\frac{\partial \mathcal{P}_{\rm g}}{\partial r}\right]\frac{x^2-y^2}{r^2}\\
\mathcal{P}_{\rm U}(x,y)
&=\frac{U_{\rm src}}{I_{\rm src}}\left\{\mathcal{P}_{\rm g}(r)+\frac{\sigma_{\rm f}^2}{4}\left[\frac{\partial^{2} \mathcal{P}_{\rm g}}{\partial r^{2}}+\frac{1}{r}\frac{\partial \mathcal{P}_{\rm g}}{\partial r}\right]\right\}+ \frac{\sigma_{\rm f}^2}{4}\left[\frac{\partial^{2} \mathcal{P}_{\rm g}}{\partial r^{2}}-\frac{1}{r}\frac{\partial \mathcal{P}_{\rm g}}{\partial r}\right]\frac{2xy}{r^2}
\end{align}

\subsubsection{Images of point and extended sources}
\label{sec:summary}

\begin{figure}[H]
    \centering
    \includegraphics[width=16cm, bb = 60 60 700 290, clip]{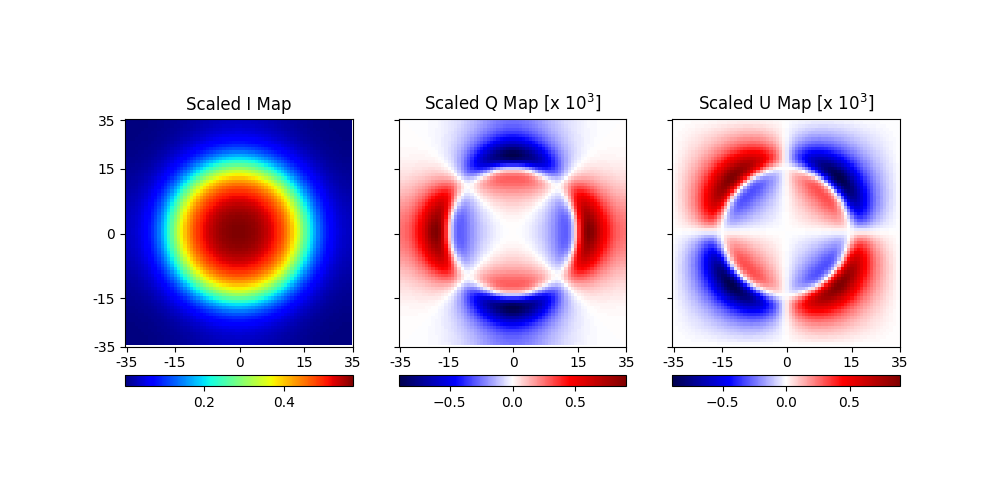}\\
    \includegraphics[width=16cm, bb = 60 60 700 290, clip]{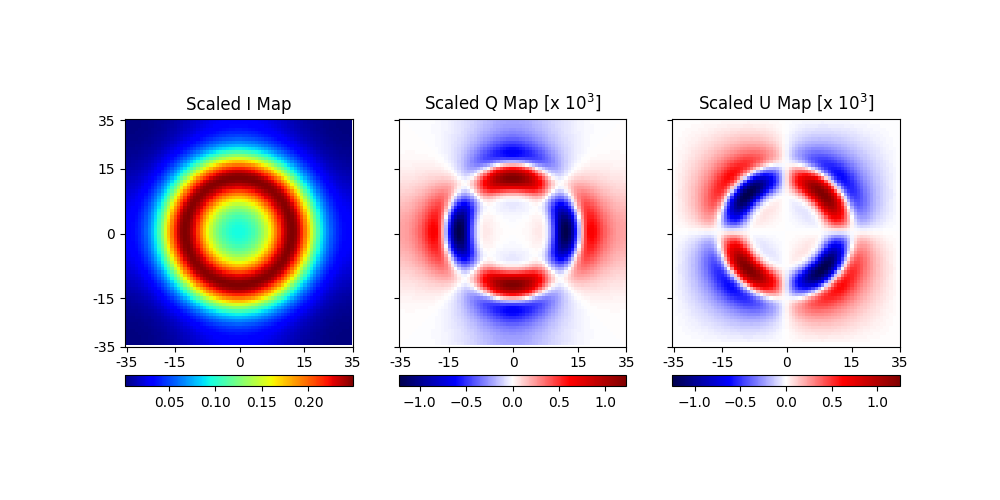}
    \caption{Upper panel: Intensity (left), Stokes Q (center) and U(right) maps, for a uniform unpolarized disk of unitary surface brightness and radius equal to 15~arcses, assuming for the geometrical PSF the one given by \citet{Fabiani_Costa+14a} at 2.95~keV , and taking $\sigma_{\rm f} = 1$ arcsec (recall that Q and U scales as $\sigma_{\rm f}^2$). Axis are in arcsec, reference frame is chosen such that Q is positive for polarization aligned along the $x$-axis. Bottom panel: the same as the top but for a uniform unpolarized ring with outer radius equal to 15~arcses, and inner one equal to 10~arcsec.}
    \label{fig:fig2}
\end{figure}

The effect of polarization-shifting can be formalized introducing different PSFs for each Stokes parameters $I_{\rm src}$, $Q_{\rm src}$ and $U_{\rm src}$ of a point source located at the coordinates origin, according to the following rules, to the first order in the  variance $\sigma_{\rm f}^2$ of the  displacement of the absorption point due to reconstruction errors.\\
\\
Its image will have an Intensity given by:
\begin{align}
I(x,y) = I_{\rm src}\left\{\mathcal{P}_{\rm g}(r)+\frac{\sigma_{\rm f}^2}{4}\left[\frac{\partial^{2} \mathcal{P}_{\rm g}}{\partial r^{2}}+\frac{1}{r}\frac{\partial \mathcal{P}_{\rm g}}{\partial r}\right]\right\} +\frac{\sigma_{\rm f}^2}{8}\left[\frac{\partial^{2} \mathcal{P}_{\rm g}}{\partial r^{2}}-\frac{1}{r}\frac{\partial \mathcal{P}_{\rm g}}{\partial r}\right]  \left[Q_{\rm src}\frac{x^2-y^2}{r^2}+U_{\rm src}\frac{2xy}{r^2} \right] \label{eq:ilin}
\end{align}
and Stokes Q-U maps according to:
\begin{align}
Q(x,y)
&=Q_{\rm src}\left\{\mathcal{P}_{\rm g}(r)+\frac{\sigma_{\rm f}^2}{4}\left[\frac{\partial^{2} \mathcal{P}_{\rm g}}{\partial r^{2}}+\frac{1}{r}\frac{\partial \mathcal{P}_{\rm g}}{\partial r}\right]\right\}+ I_{\rm src}\frac{\sigma_{\rm f}^2}{4}\left[\frac{\partial^{2} \mathcal{P}_{\rm g}}{\partial r^{2}}-\frac{1}{r}\frac{\partial \mathcal{P}_{\rm g}}{\partial r}\right]\frac{x^2-y^2}{r^2}\label{eq:qlin}\\
U(x,y)
&=U_{\rm src}\left\{\mathcal{P}_{\rm g}(r)+\frac{\sigma_{\rm f}^2}{4}\left[\frac{\partial^{2} \mathcal{P}_{\rm g}}{\partial r^{2}}+\frac{1}{r}\frac{\partial \mathcal{P}_{\rm g}}{\partial r}\right]\right\}+ I_{\rm src}\frac{\sigma_{\rm f}^2}{4}\left[\frac{\partial^{2} \mathcal{P}_{\rm g}}{\partial r^{2}}-\frac{1}{r}\frac{\partial \mathcal{P}_{\rm g}}{\partial r}\right]\frac{2xy}{r^2}\label{eq:ulin}
\end{align}
The fact that different Stokes parameters have different PSFs, is akin of what happens in chromatic aberration, where radiations at different wavelengths have different PSFs, and for this reason this phenomena can be thought of as \textit{Stokes chromatism}. The fact that it manifests itself on calibration as an excess radial polarization, is just a consequence of the geometry of the source, and of the profile of the geometrical PSF. \\
\\
As an example of the possible polarization patterns that this effect can produce in Fig.~\ref{fig:fig2} we show the results in the case of a uniform disk, and a uniform ring adopting for simplicity as geometrical PSF the Gaussian-King model given by \citet{Fabiani_Costa+14a} at 2.95~keV.

\subsubsection{Limits of the Linearized Approach}
\label{sec:limit}
The validity of the linearized approach is limited by two conditions: $\sigma_{\rm f} \ll \mathcal{P}_{\rm g}/|\nabla \mathcal{P}_{\rm g}|$, and $\sigma_{\rm f}$ must be small enough to allow linearization in transforming from cartesian to polar coordinartes (see Eq.~\ref{eq:eq6}-\ref{eq:eq7}). Both conditions are likely to fail first in the origin, where the geometrical PSF has a sharp peak. Indeed any centrally peaked geometrical PSF can be written as:
\begin{align}
\mathcal{P}_{\rm g}(r)  = \sum_{i=0}^\infty a_i r^{2i} \quad{\rm with}\quad a_1<0
\end{align}  
which means that the correction to the PSF, in the very center, due to the displacement of the absorption point at the linear order is $a_1\sigma_{\rm f}^2 <0$. Obviously for
$\sigma_{\rm f}^2 < \mathcal{P}_{\rm g}(0)  / (\partial^{2} \mathcal{P}_{\rm g}/\partial r^{2})_{r=0}$ the resulting total PSF $\mathcal{P}_{\rm I}(0)  $ becomes negative in the origin. But even at smaller values of $\sigma_{\rm f}$, the linear correction can be such that the  resulting total PSF display a minimum in the origin, which can never happen in a full convolution process. In general we expect that as  $\sigma_{\rm f}$ increases the error associated with the linear approach will also progressively increase, mainly in the central region at $r< $ a few $\sigma_{\rm f}$.  The exact threshold for the validity of the linear approach will depend on the shape of the geometrical PSF  $\mathcal{P}_{\rm g}(r) $ as well as on the distribution function describing the chance that the reconstructed absorption point is misplaced.

We will assess here this limit adopting the V11 circular PSF from the \texttt{ixpeobssim} software package \citep{Baldini_Bucciantini+22a}, as a model for the geometrical PSF  $\mathcal{P}_{\rm g}(r) $. This PSF has been optimized to reproduce the results of the IXPE satellite, still retaining the circular symmetry, and as such it provides the best starting point for the application of our formalism to true observational data. Caveats  with this choice will be discussed later. The probability distribution function of the spurious displacement due to reconstruction errors, is assumed to be a one-dimensional Gaussian with variance  $\sigma_{\rm f}$.  We can then perform numerically the full convolution process of Eq.~\ref{eq:1}, and compute the related PSF for each Stokes parameter, to be compared with the linearized results. We will consider the intensity distribution I, and the polarized fraction distribution $\sqrt{Q^2+U^2}/I$, which, for an unpolarized sources, are just a function of the radial distance. For highly polarized source images will be elongated, an one will have to consider the full image pattern. 

\begin{figure}[H]
    \centering
    \includegraphics[width=6cm]{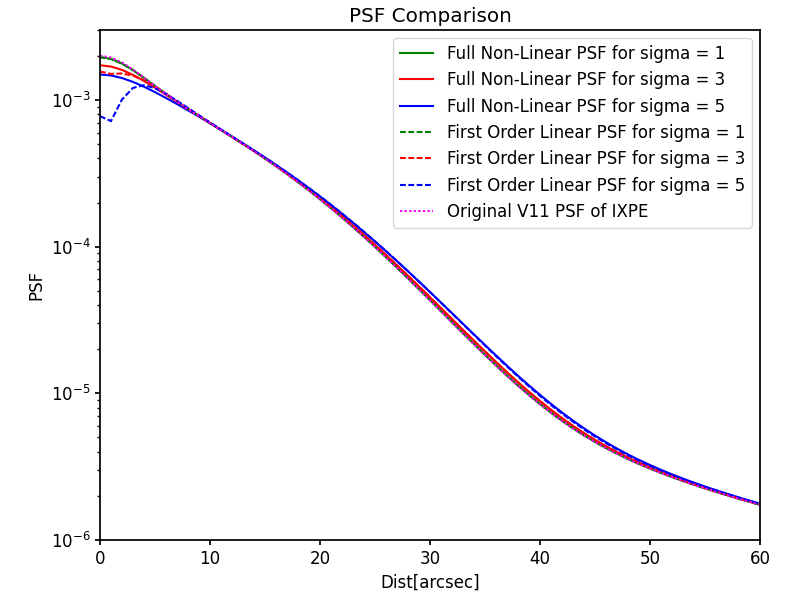}\includegraphics[width=6cm]{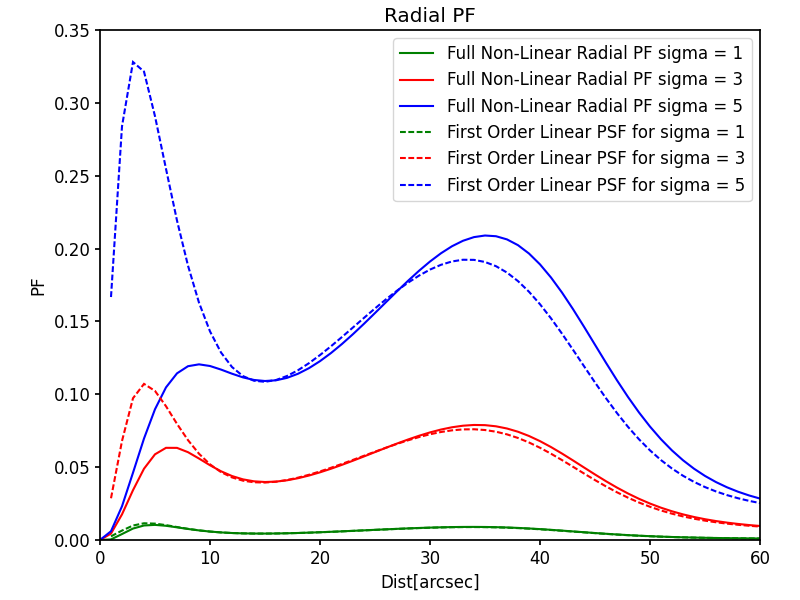}\includegraphics[width=6cm]{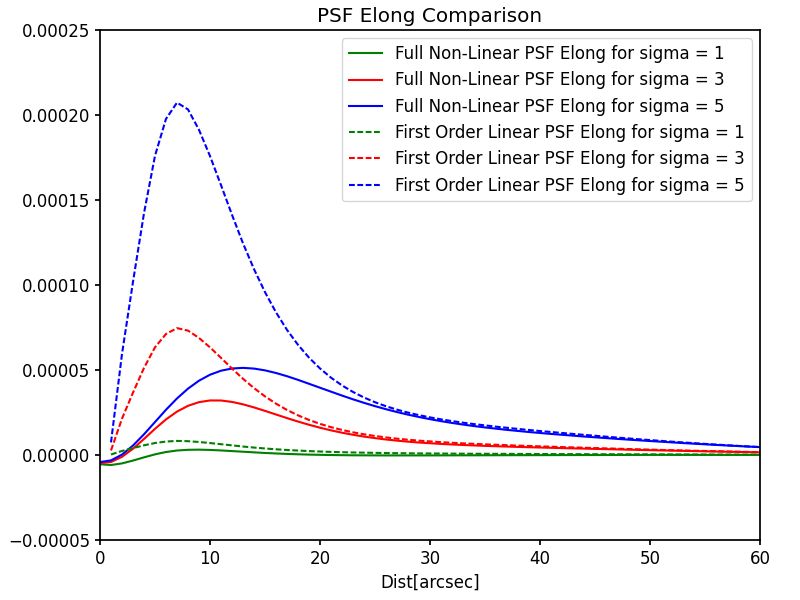}
    \caption{Left panel: comparision of the PSF computed with the linearized approach and fully nonlinear approach for various values of $\sigma_{\rm f}$ for an unpolarized point source. Central panel:  comparison of the radial polarization as a function of distance from the source position  computed with the linearized approach and fully nonlinear approach for the same values of $\sigma_{\rm f}$. Right panel: elongation of the PSF for a fully (100\%) polarized point source computed with the linearized approach and fully nonlinear approach. In all cases $\mathcal{P}_{\rm g}(r)$ is given by the V11 PSF of \texttt{ixpeobssim}, and $\mathcal{P}_{\rm f}$ is assumed to be a one-dimensional Gaussian.}
    \label{fig:fig3}
  \end{figure}

In Fig.~\ref{fig:fig3} we compare the total PSF of the intensity I, for $\sigma_{\rm f} = {1,3,5}$~arcsec, together with the polarized fraction as a function of the  distance fom the origin, where the point source is supposedly located. Variations of the total PSF for I are small, mostly concentrated either in the origin, where, for $\sigma_{\rm f} = 3$~arcsec, we begin to observe, for the PSF computed in the linearized regime,  the profile inversion we discussed before, or in the low intensity wings at $r>50$~arcsec. We can safely claim that for $\sigma_{\rm f} \le 3$~arcsec, the \effect~ leads to marginal changes in the intensity distribution of unpolarized sources. The linearized approach however tends to lead to an eccessive suppression of the intensity in the central region which has important implications in the estimate of the polarized fraction.  By looking at the polarized fraction distribution we see that for  $\sigma_{\rm f} \le 3$~arcsec, the linearized approach matches quite well the results of the full non linear convolution. For  $\sigma_{\rm f} > 3$~arcsec the discrepancy in the central region, $r< 5$~arcsec, becomes progressively stronger. In the linearized approach a double peak structure emerges, that is partly present also in the full convolution, but  showing an inner peak that grows much more, and that is in large part associated to the error in the intensity profile. There is a much better agreement at $r>10-12$~arcsec for $\sigma_{\rm f}$ up to $5$~arcsec. For higher values of $\sigma_{\rm f}$, even beyond 15 arcsec, the agreement is quite poor: at $\sigma_{\rm f}=7$~arcsec, that maximum polarized fraction with the full convolution is found to be $0.38$ at 36~arcsec, while in the linearized approach is $0.33$ at 33~arcsec. This however is not surprising given that at  $\sigma_{\rm f}=7$~arcsec the displacement due to error reconstruction becomes comparable with the one due to the geometric PSF.

In Fig.~\ref{fig:fig3}  we also show the elongation of the image, for a source with 100\% polarization, defined as the difference between the intensity profile in the direction aligned with the polarization of the source, with respect to the intensity of an unpolarized source. Interestingly even in the full non linear case the elongation is correctly described by a $\cos{(2\phi_1)}$  pattern. It is just the radial profile that changes with $\sigma_{\rm f}$. We see that similarly to what happen for the PSF of unpolarized sources, the linear approximation tends to overestimate the \effect~  at smaller radii, while converging to the non linear results at $r>20$~arcsec. In general however this effect is quite small, likely to be appreciable only for larger values of  $\sigma_{\rm f}$, and only for highly polarized sources. Given the typical polarization of sky sources $< 20\%$ this effect can safely be neglected.


\section{Comparison with data and GPD simulations}
\label{sec:comp}

We are going here to compare our simple formalism, in the full non-linear regime, with both the results of a full GPD simulation using the \texttt{GEANT4} simulator \citep{Agostinelli_Allison+03a}, and of an IXPE observation of a bright unpolarized point source (Cygnus X-1). We do not expect our simplified approach to exactly match the true response of the instrument, but this comparison offers us a way to better characterize its strengths and weaknesses.  For the \texttt{GEANT4} simulation we considered a point source, with a photon distribution described by a simple power-law with photon index $-2$. The \texttt{GEANT4} simulator is coupled with the \texttt{ixpeobssim/ixpesim} software \citep{Baldini_Bucciantini+22a,Di_Lalla19}\footnote{for \texttt{ixpesim} see  https://etd.adm.unipi.it/theses/available/etd-04042019-100412/}, that provides the optical response of the IXPE satellite, using the V11  PSF for $\mathcal{P}_{\rm g}(r)$. This approach is, at this time, the best available in order to simulate the true response of the IXPE satellite. There are however a few important caveats in this approach: the V11 PSF is circular, while the IXPE calibration shows that the various mirror modules have PSFs with significant non circular features; the  V11 PSF is optimized to respoduce the circular average intensity of an IXPE point source within its half power diameter, and as such it already factors in the widening of the PSF due to the \effect. Using the V11 PSF amounts to double counting the \effect. The \texttt{GEANT4} simulation and the IXPE observation are reduced using \texttt{ixpeobssim} with the same pipeline. The energy band band considered here is the operating IXPE band $[2-8]$~keV. 

\begin{figure}[H]
    \centering
    \includegraphics[width=8cm]{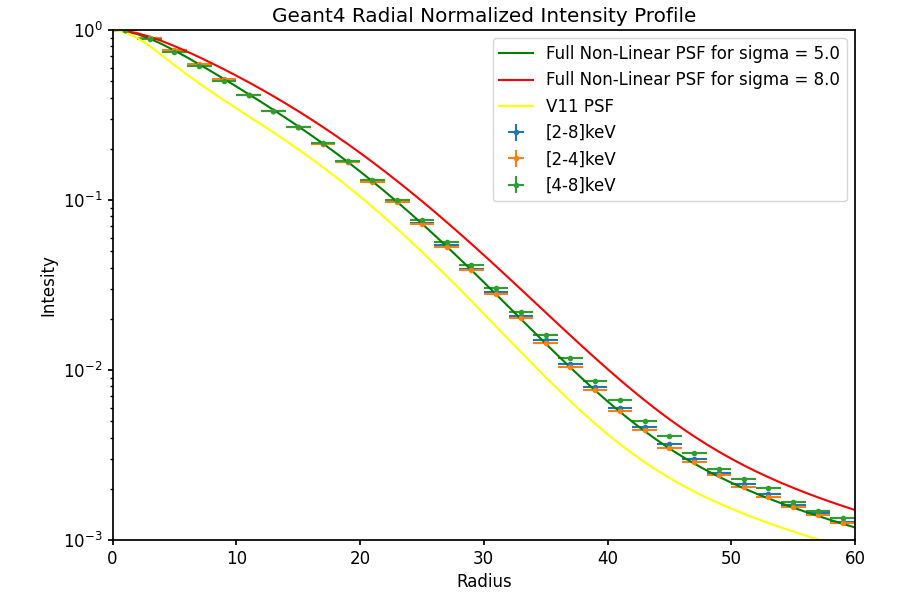}\includegraphics[width=8cm]{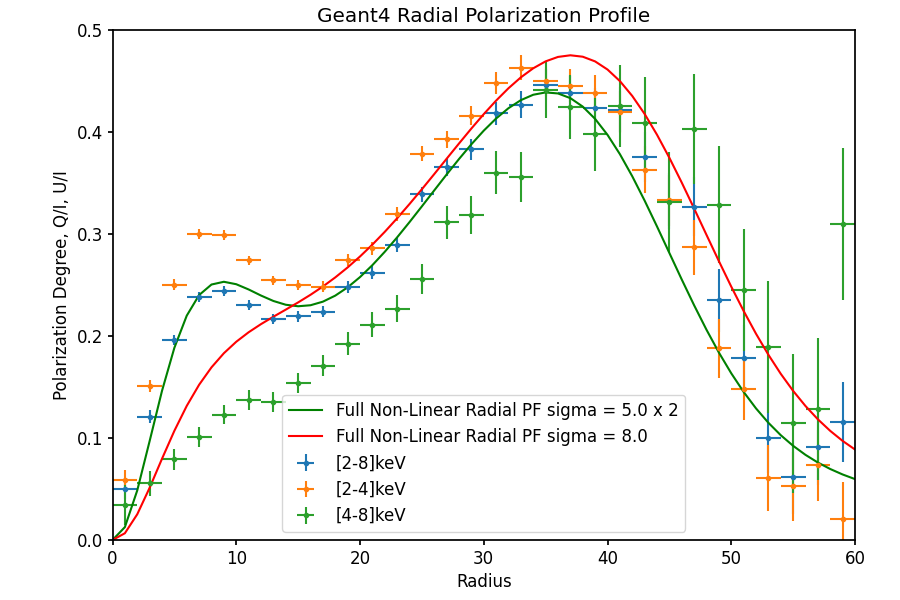}\\
        \includegraphics[width=8cm]{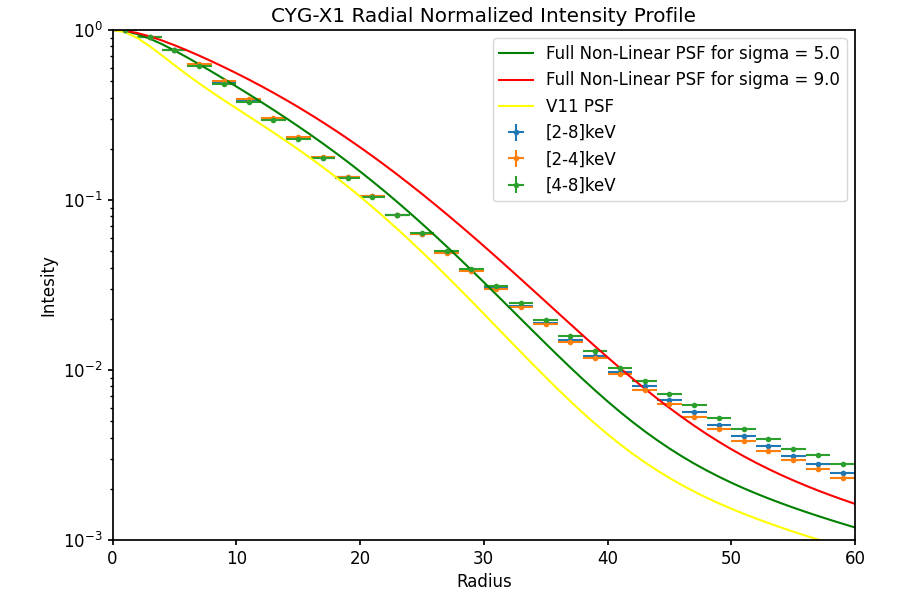}\includegraphics[width=8cm]{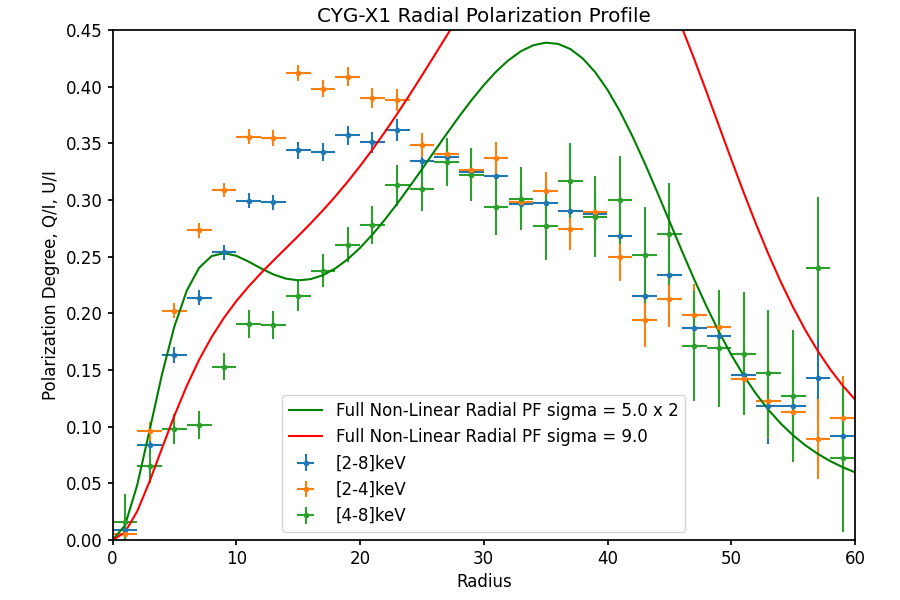}\\
        \caption{Upper left panel: Intensity profile (in various energy bands) of a point source, with photon index $=-2$ simulated with the \texttt{GEANT4}, and assuming the V11 \texttt{ixpeobssim} PSF as input geometrical PSF, compared both with the V11 PSF itself, and with the intensity profile computed with our fiducial non-linear formalism for two different values of the variance $\sigma_{\rm f}$. Upper right panel: comparison of the profile of radial polarization  (in various energy bands between the same \texttt{GEANT4} simulation and our fiducial non-linear formalism (the $\sigma_{\rm f}=5$ case has been multiplied by a factor 2). Lower panels: the same as the above ones but for Cygnus X-1 as observed by the IXPE satellite. }
    \label{fig:fig4}
  \end{figure}

In Fig.~\ref{fig:fig4} we show a comparison of the  \texttt{GEANT4} simulation with our formalism, as described in Sect.~\ref{sec:limit} where we have again assumed that the probability distribution function of the spurious displacement due to reconstruction errors, is a one-dimensional Gaussian with variance  $\sigma_{\rm f}$.  By looking at the profiles from the \texttt{GEANT4} simulation, we see that the effect of \effect~  on the PSF is clearly present, with the wings of the PSF typically twice brighter than expected. However the dependence on the energy is small. The amount of radial polarization is instead quite large, up to 45\%. Interestingly there is a clear dependence of the radial polarization profile on the selected energy range, with a double peak structure in the lower energy range, not present at higher energies. This is most likely an indication that  the probability distribution function of the spurious displacement does not have the same functional form at all energies (as we assume in our formalism when we adopt a Gaussian where the only parameter that one can adjust is the variance).

If we model the \effect~  using our simplified non-linear formalism, we get results that, in terms of radial polarization profile, are in fair agreement with the \texttt{GEANT4} simulation, if we assume $\sigma_{\rm f}=8$~arcsec. Both the amont of polarization and the location of the peak between 30 and 40~arcsec, are recovered. However as discussed above, we cannot reproduce the change of the polarization profile with energy, and we do not get the inner peak. Moreover, by looking at the intensity profile we notice that results using $\sigma_{\rm f}=8$~arcsec tend to over-predict by about a factor 2 the widening of the PSF. This is not unexpeced, given that the probability distribution function of the spurious displacement is not one dimensional (it might even not be centrally peaked), which breaks the one-to-one relation between the level of radial polarization and the PSF widening. We noticed that we can get a much better agreement of the PSF if we set $\sigma_{\rm f}=5$~arcsec, which however under-predict by a factor 2 the level of radial polarization. It appears that the \texttt{GEANT4} simulation is well reproduced, on average in the full energy range, if we model the \effect~  with $\sigma_{\rm f}=5$~arcsec, but then increase by a factor 2 the amount of radial polarization. In this case both the intensity and polarization profile are recovered. This will be referred in the following as our \textit{fiducial} non-linear approach/formalism.

In Fig.~\ref{fig:fig4} we show the radial profile, averaged over circular annuli, of the intensity and radial polarization observed by IXPE for Cygnus X-1. Ideally one would like to select a very bright and unpolarized source. In practice, however, a low level of intrinsic polarization (as long as it is small enough to neglect higher order effect, like image distortion), is not a problem, given that, once averaged over an annulus, its contribution to the radial polarization pattern vanishes (the polarized fraction of Cygnus X-1 is just a few percents). By looking at the intensity profile one can see immediately that the results are not in agreement with either the V11 PSF or the \texttt{GEANT4} simulations. In particular the   \texttt{GEANT4} simulation appears to over-predict by about 20\% the intensity in the central region at $r<30$~arcsec, and to greatly under-predict it at larger distances, up to a factor 2. This is a clear indication that the V11 PSF that is used in the \texttt{GEANT4} simulation, despite having been optimized on calibration data, does not allow us to reproduce the true intensity profile. Part of the problem if due to the fact that the V11 was calibrated on data already affected by \effect, and as such its use in a \texttt{GEANT4} simulation ends up double counting it. This is likely the origin of the over-prediction in the central region. The fact that the wings of the PSF at $r>30$~arcsec, are under-predicted, is most likely due to either poor calibration in this low intensity region, or possible differences between calibration data and flight performances. More significative are the differences in the profiles of radial polarization. The maximum amount of radial polarization is $\sim 40\%$, similar to what is found in the \texttt{GEANT4} simulation, however, in Cygnus X-1 there is no evidence, of any double peaked structure, and the radial polarization tend to peak much closer to the source: at 20~arcsec in the [2-4]~keV, and at 30~arcsec in the [2-8]~keV energy bands. Part of this difference might be attributed to the already discussed difference in the intensity profiles, due to the incorrect use of the V11 PSF.

We have tried to see if it was possible to modify the geometrical PSF $\mathcal{P}_{\rm g}(r)$ with respect to the V11 in order to reproduce both the intensity  and radial polarization profiles. We did that with our fiducial non-linear formalism, that has proved to approximate in a reliable way the results of the \texttt{GEANT4} simulation, but could not find a satisfactory solution. It is easy to modify the geometrical PSF, to match the observation, but we could not recover the radial polarization profile. This could be due to an intrinsic limitation of our formalism (e.g. the assumption that the probability distribution function of the spurious displacement due to reconstruction errors, is a one-dimensional Gaussian), but we cannot exclude that the presence of non-circular patterns in the IXPE PSF might somehow be important. There is also the chance that the \texttt{GEANT4} simulation of an idealized GPD, does not fully capture the reality of the true GPD on board of the satellite. This is not unexpected given that we already know of spurious effects that cannot be reproduced in the \texttt{GEANT4} \citep{Rankin_muler+22a}.

While the \texttt{GEANT4} simulation, and our formalism get consistent results, there are still significative differences with respect to real IXPE observations of point sources.  However, as we will show in the next section, for real extended sources, this effect is only mildly dependent on the details of the radial profile for a point source, making this approach more robust that what could be guessed from just the simple point source analysis.

\section{Mueller Matrix Formalism}
\label{sec:mueller}

The generalization of the PSF for polarized observations, is the so called Mueller Matrix $\mathcal{M}$ \citep{Tinbergen96a}, whose elements represent the images of a polarized point source in the Stokes plane, as a function of its intrinsic Stokes parameters: $I_{\rm src}$, $Q_{\rm src}$, and $U_{\rm src}$.  Its diagonal elements describe how each Stokes parameter of the source is turned into an image in the same Stokes parameter, while off-diagonal elements describe polarization leakage from one Stokes parameter to another.

For example the terms in curly brakets in Eq.s~\ref{eq:ilin}-\ref{eq:ulin}, represent the diagonal elements $\mathcal{M}_{II}$, $\mathcal{M}_{QQ}$ and $\mathcal{M}_{UU}$ respectively, while the terms in square brakets represent the off-diagonal terms $\mathcal{M}_{IQ}$, $\mathcal{M}_{IU}$, $\mathcal{M}_{UI}$ and  $\mathcal{M}_{QI}$. It can be shown that the spurious displacement of the reconstructed absorption point due to reconstruction errors, does not cause any mixing between U and Q ($\mathcal{M}_{QU}=0$). Moreover it can easily be shown that in our case $\mathcal{M}_{IQ} = \mathcal{M}_{IU}$, and $\mathcal{M}_{UI}= \mathcal{M}_{QI}$.

As long as the geometrical PSF $\mathcal{P}_{\rm g}(r)$ is circular Eq.s~\ref{eq:ilin}-\ref{eq:ulin}, can be generalized to:
\begin{align}
I(x,y) &= I_{\rm src} \mathcal{G}(r) +  \mathcal{F}(r)\left[Q_{\rm src}\frac{x^2-y^2}{r^2}+U_{\rm src}\frac{2xy}{r^2} \right]\label{eq:mmatrix1}\\
Q(x,y) 
  &=Q_{\rm src}\mathcal{H}(r) + I_{\rm src}
   \mathcal{K}(r) \frac{x^2-y^2}{r^2}\label{eq:mmatrix2}\\
U(x,y)
&=U_{\rm src}\mathcal{H}(r) + I_{\rm src}\mathcal{K}(r) \frac{2xy}{r^2}\label{eq:mmatrix3}
\end{align}
where $\mathcal{G}(r), \mathcal{F}(r), \mathcal{H}(r), \mathcal{K}(r)$ are the Mueller matrix element. Determination of these functions allows one to compute the image of any source (polarized or un polarized) in each Stokes parameter. In Sect.~\ref{sec:linear} we have shown how to derive these functions from the optical PSF at first order in the variance of  the probability distribution function of the spurious displacement due to reconstruction errors, which provides a good approximation as long as the maximum amount of radial polarization does not exceed 15\%. For larger values one need to use the full non linear approach. It is also possible to derive those functions by fitting directly either the results of a full \texttt{GEANT4} simulation, or even the observation of a bright point source.  We have shown that, in the case of IXPE, the non linear approach based on the assumption that the probability distribution function of the spurious displacement due to reconstruction errors is a one-dimensional Gaussian, can be tuned easily to reproduce very accurately the result of a \texttt{GEANT4} simulation. So we expecxt these two approaches to lead to equivalent results. On the other hand we have noticed important differences in comparison with a true observation of Cygnus X-1.

Our investigation have shown that $\mathcal{G}(r) = \mathcal{H}(r)$ and that $\mathcal{F}(r)$ can be safely neglected, given that its contribution to the source elongtion is small, and typically the intrinsic polarization of X-ray sources does not exceed 20\%. On the other hand  a correct determination of $\mathcal{K}(r)$ is quite important because it can easily lead to false detection. In general this is not an issue for isolated point sources, but in extended sources it can lead to complex polarization patterns, that in case of small intrinsic polarization, can dominate over the true one.

A typical way to evaluate this effect, for IXPE observation,  is to simulate a full IXPE observation of an extended source by coupling the \texttt{ixpeobssim/ixpesim} software, which provide the optical response of the various mirror modules, with a  \texttt{GEANT4} simulation that gives the response of the GPD, processing the resulting tracks with the same data reduction software that is used on the true date, to get Level 2 files that are then analysed again as true data are. This however can be extremely time consuming, in general requires a huge statistics in order to keep Poisson counting noise under control, and still suffers from the limitations discussed in Sect.~\ref{sec:comp} (formost $\mathcal{P}_{\rm g}(r)$ is not known).

\begin{figure}[H]
    \centering
    \includegraphics[width=15cm, bb = 60 50 700 320, clip]{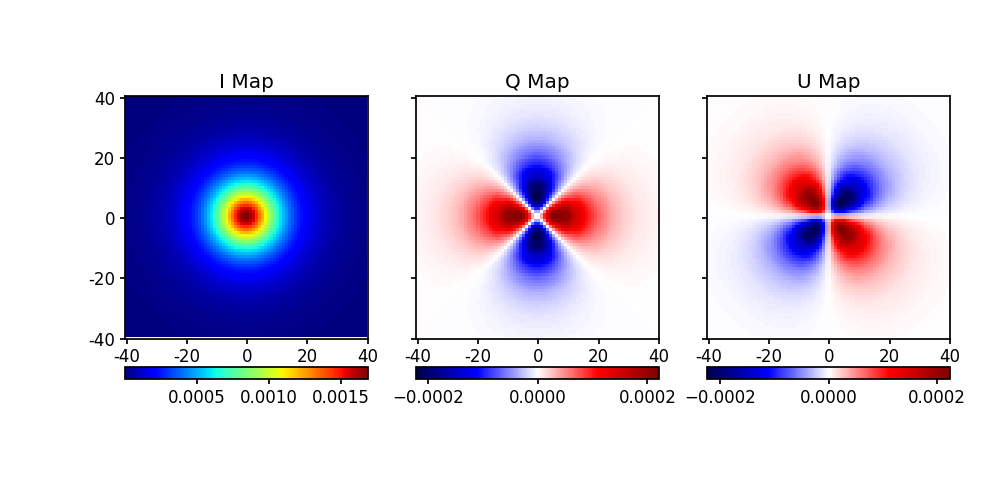}\\
        \includegraphics[width=15cm, bb = 60 50 700 320, clip]{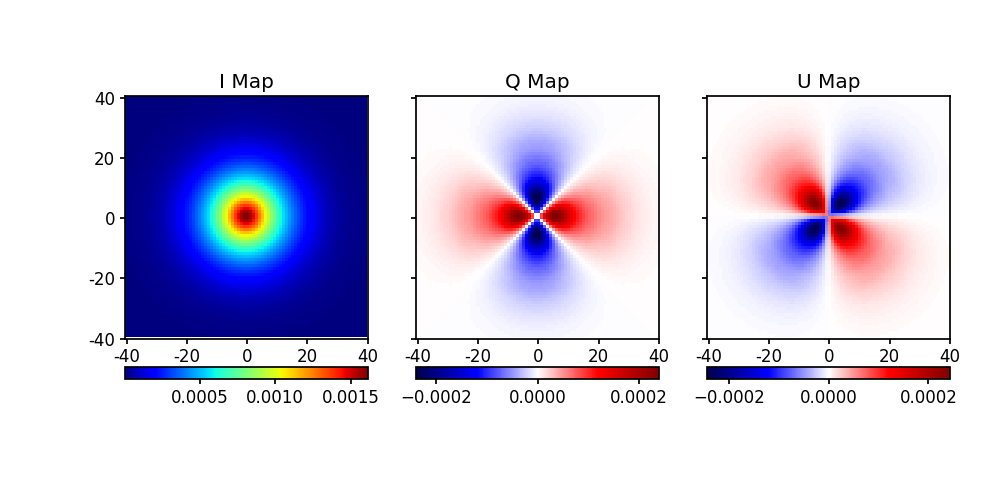}\\
        \caption{Upper panels: Maps of I, U and Q for a point source computed according to Eq.S~\ref{eq:mmatrix1}-\ref{eq:mmatrix3} with functions $\mathcal{G}(r)$ and $\mathcal{K}(r)$, fitted on a \texttt{GEANT4} simulation of an upolarized point source.  Lower panels: the same as the above ones but for functions $\mathcal{G}(r)$ and $\mathcal{K}(r)$, fitted on the Cygnus X-1 observation by the IXPE satellite.  The reference frame is chosen such that Q is positive for polarization aligned along the $x$-axis.}
    \label{fig:fig5}
  \end{figure}

A much easier approach would be to derive the functions $\mathcal{G}(r)$ and $\mathcal{K}(r)$, either from a \texttt{GEANT4} simulation, or from true on-flight data, and use them to compute the Mueller matrix elements. Then computing a polarization map of any source, reduces to a simple  and numerically inexpensive convolution. In Fig.~\ref{fig:fig5} we show the Mueller matrix elements (of an unpolarized source), obtained by fitting  $\mathcal{G}(r)$ and $\mathcal{K}(r)$ to the \texttt{GEANT4} simulation results (but we get equivalent maps using our  fiducial non-linear formalism), and to the IXPE observation of Cygnus X-1. There are some differences, with the true data giving Q and U maps  more peaked toward the center.

\begin{figure}[H]
    \centering
    \includegraphics[width=6.2cm, bb = 10 0 500 330, clip]{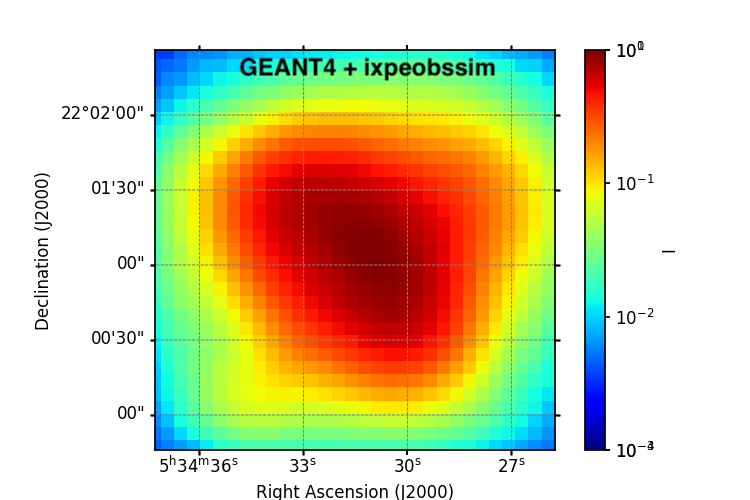}\includegraphics[width=5cm, bb = 105 0 500 330, clip]{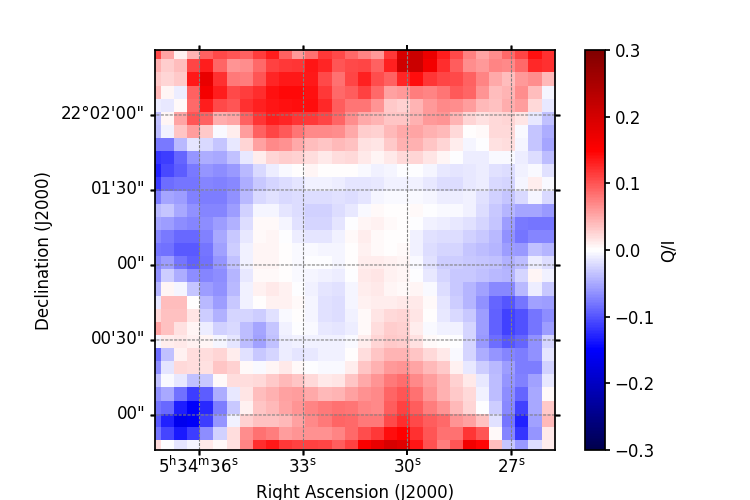}\includegraphics[width=5cm, bb = 105 0 500 330, clip]{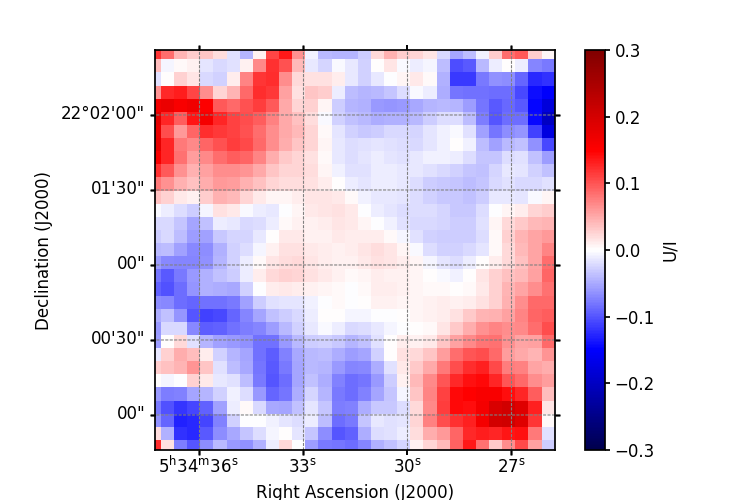}\\
    \includegraphics[width=6.2cm, bb = 10 0 500 330, clip]{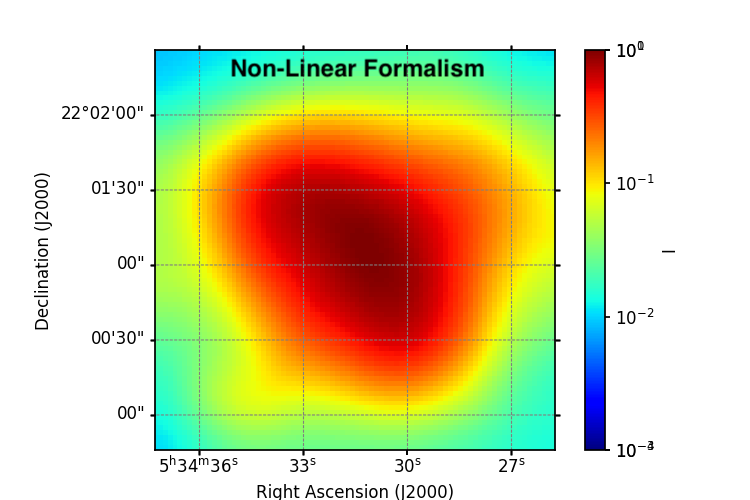}\includegraphics[width=5cm, bb = 105 0 500 330, clip]{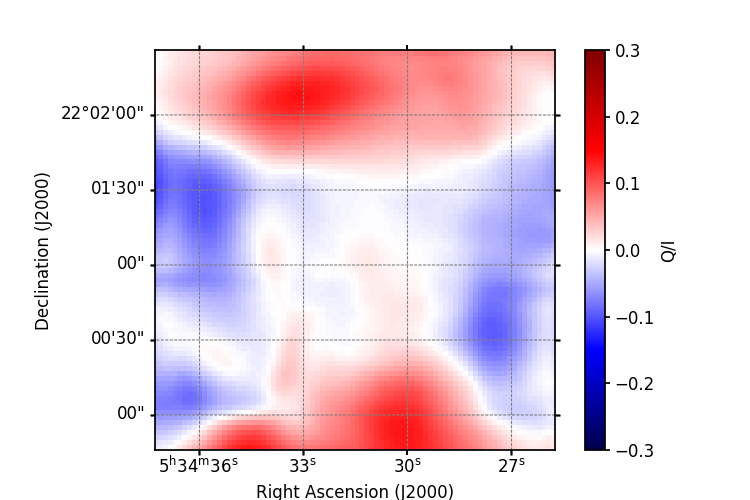}\includegraphics[width=5cm, bb = 105 0 500 330, clip]{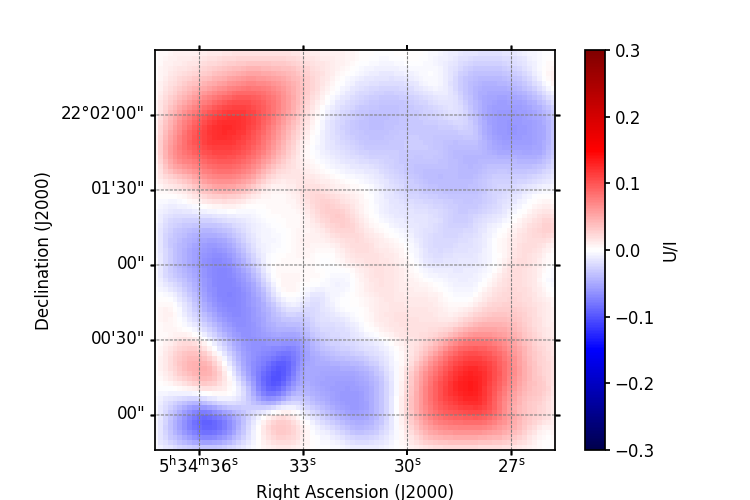}\\
    \includegraphics[width=6.2cm, bb = 10 0 500 330, clip]{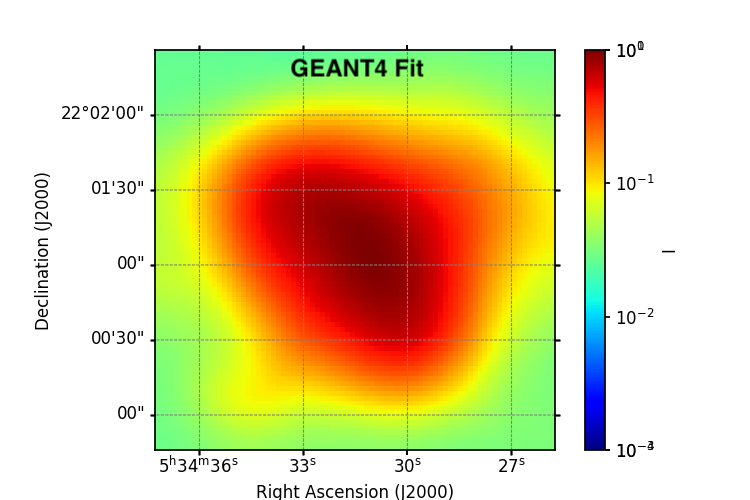}\includegraphics[width=5cm, bb = 105 0 500 330, clip]{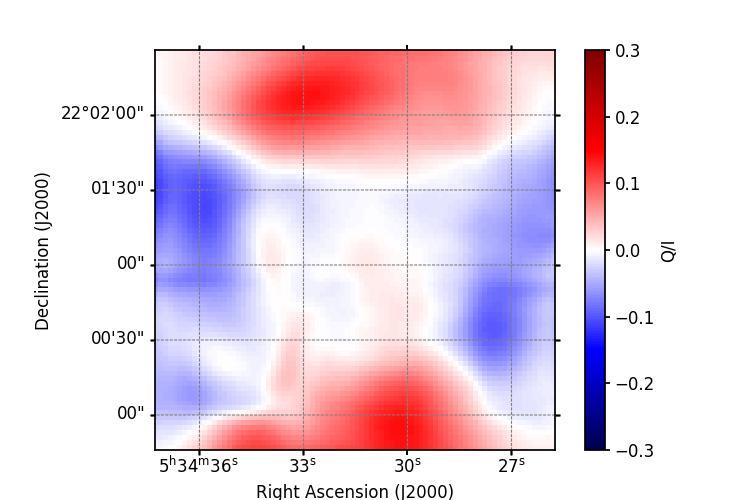}\includegraphics[width=5cm, bb = 105 0 500 330, clip]{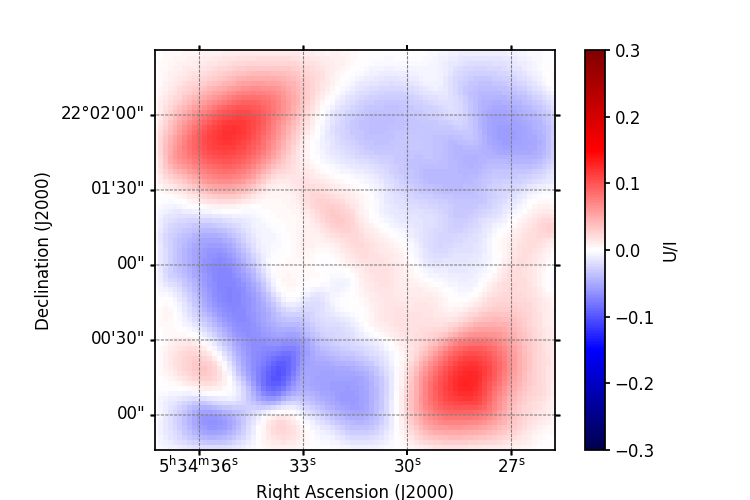}\\
                \includegraphics[width=6.2cm, bb = 10 0 500 330, clip]{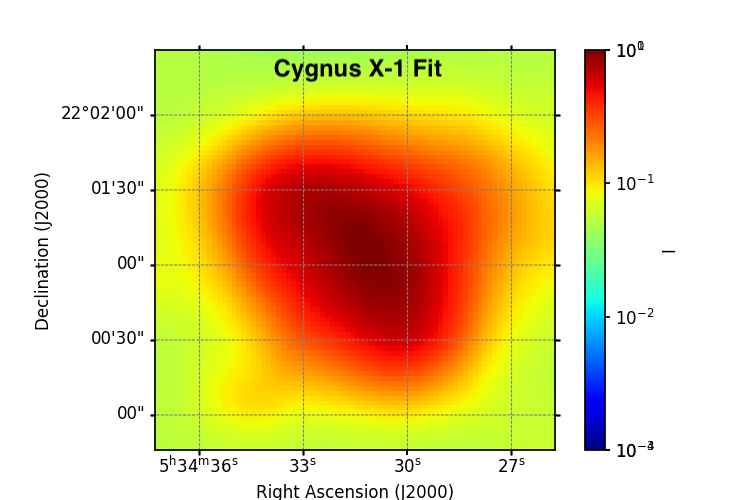}\includegraphics[width=5cm, bb = 105 0 500 330, clip]{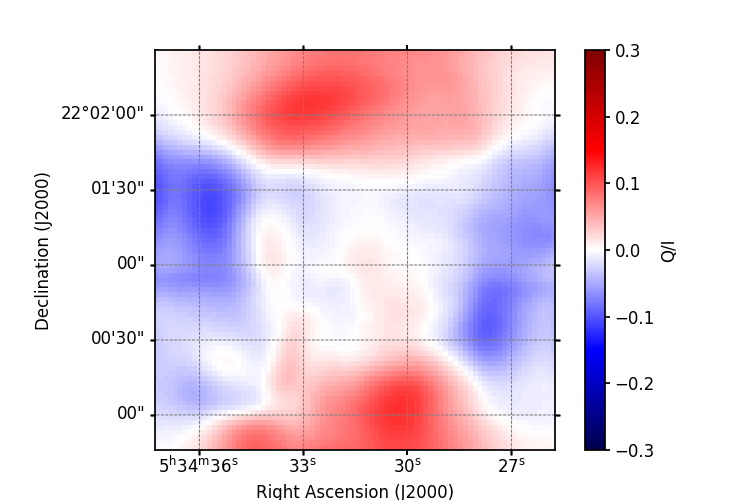}\includegraphics[width=5cm, bb = 105 0 500 330, clip]{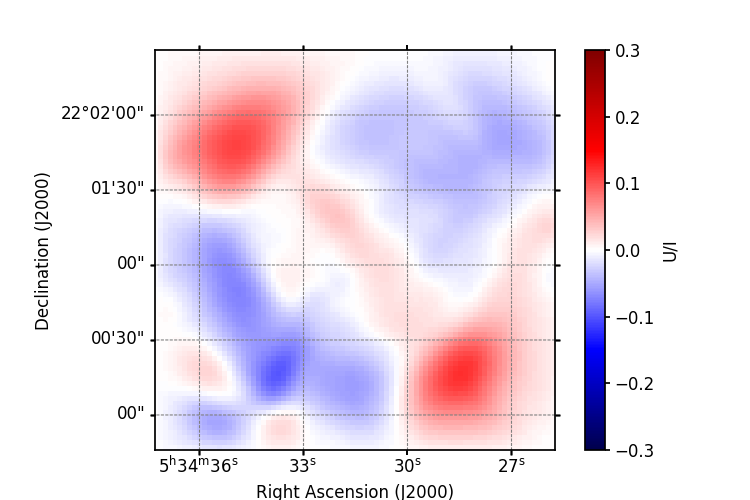}\\
        \caption{Upper panlel: Image in I, Q/I and U/I of the Crab nebula in the [2-8]~keV band assuming fully unpolarized radiation, and computed using a full  \texttt{ixpeobssim} plus \texttt{GEANT4} simulation. Data where reduces using standard \texttt{ixpeobssim} data analysis tool. Gaussian smoothing was applied to reduce Poisson noise. The orignal source input file was the Chandra ObsId 16364. Lower rows: the same as the upper one, but computed by convolution of the Mueller matrix element with the intensity map of  Chandra ObsId. In order from the second to the fourth row, the functions $\mathcal{G}(r)$ and $\mathcal{K}(r)$ where derived as:  non-linear formalism (second row), fit of the results from a \texttt{GEANT4} point source simulation (third row),  fit of the Cygnus X-1 data  (forth row).  The reference frame is chosen such that Q is positive for polarization aligned along the N-S direction.}
    \label{fig:fig6}
  \end{figure}

In Fig.~\ref{fig:fig6} we show the application of the Mueller matrix formalism with the functions $\mathcal{G}(r)$ and $\mathcal{K}(r)$ computed using different approaches, to the case of a real extended source, the Crab nebula. We took a map of the intensity in the [2-8]~keV band derived from a  Chandra observation (OsbId 16364) as our fiducial intrinsic source, and we assumed the source to be fully unpolarized (not true for the Crab nebula) to see what kind of polarization pattern will arise from \effect. We also performed a full \texttt{ixpeobssim/ixpesim} plus \texttt{GEANT4} simulation, which is also shown for comparison. Results obtained with $\mathcal{G}(r)$ and $\mathcal{K}(r)$ derives either by fitting the \texttt{GEANT4} simulation of an unpolarized source, or using our fiducial non-linear formalism agree very well between them and with the full  \texttt{ixpeobssim} plus \texttt{GEANT4} simulation, showing that they provide a reliable substitute of the former, more cumbersome, approach.  Fig.~\ref{fig:fig6} shows clearly that despite the evident difference in the radial polarization profile of an unpolarixed point source, between the \texttt{GEANT4} simulation and the Cygnus X-1 data, once applied to extended sources with complex intensity map, the results show only very marginal differences. 

\section{Conclusions}
\label{sec:conclusion}

X-ray polarimetry based on Gas Pixel Detectors offers a new observational window that, for the first time, has enabled us  to do spatially-resolved polarimetric studies of Galactic and extra-Galactic extended sources, unveiling otherwise inaccessible properties. IXPE is the first instrument to successfully adopt this technique. As such its analysis also pioneers our understanding of this new technology, and a correct assessment of the possible issues related to GPD polarimetry is of great importance in the light of its use in future instruments like eXTP \citep{Zhang+19a} . It was indeed during its calibration phase that, for the first time, polarized halos were noticed around unpolarized calibration sources, spurring an investigation into the origin of this effect.  \\
\\
Here for the first time, we illustrate the origin of this \effect~, showing that is fundamentally related to errors in determining the photon absorption point at the track-reconstruction stage (a key part of GPD based approach), and in particular on the correlation of this error with the estimated EVPA. We developed a mathematical formalism that allows one to model the effect for generic sources, evaluated the range of validity of this model correction, and presented a comparison both with real data and more sophisticated Monte Carlo simulations. We have shown that the approach provides a reliable estimate of the effect, even if, for a proper data cleaning, a more robust Montecarlo is likely necessary.   By adopting the Mueller matrix formalism we were able to apply our results to complex extended sources. A more complex approach, capable of handling non-circular PSFs, is under developement.\\
\\
While for isolated point sources, this \effect~ can lead to radially polarized halos with a polarized fraction as high as 40\%, the effect on extended source, can be much smaller, depending  on the intensity gradients. In general for isolated point source, this effect is of no concern as long as the regions used for the analysis are sufficiently large and/or centered on the source. For extended source the level of polarization due to \effect~  depend mostly on how compact the source is (in comparison with the PSF), and on the level of the true intrinsic polarization. The worst case is that of a weak extended source close to a bright point source, in which case the latter's halo can dominate the signal -- one must then perform a careful subtraction for a valid polarization measurement. A \texttt{python} script to evaluate this effect  optimized for IXPE has been released together with the \texttt{ixpeobssim} package \citep{Baldini_Bucciantini+22a}. Note that the accuracy of the formalism is sufficient to get an estimate of the effect, but not to perform  proper data cleaning.\\
\\
Obviously the level of this problem depends on the magnitude of the position mis-localization and its correlation with the EVPA. For the default IXPE track `moments analysis' \citep{AnalyticRecon}, the effect can be quite substantial as we have shown. For other track analysis schemes, the effect will  differ for the same events. For example preliminary analysis of a neural net track measurement scheme (`NN analysis'), indicates that the effect is $\sim 3\times$ smaller, but still present \citep{TakaoCNNclassification, CNNregression, StanfordCNN, ROMANI2021}. However any reconstruction algorithm will have some inaccuracy in determining the impact point, so this effect is likely to always be present in GPD-based polarimetric observations.


\begin{acknowledgements} NB was supported by the INAF MiniGrant "PWNnumpol - Numerical Studies of Pulsar Wind Nebulae in The Light of IXPE" \end{acknowledgements}

\appendix
\section{Generalization of the Linear Formalism}
\label{sec:general}
\begin{figure}[H]
    \centering
    \includegraphics[width=10cm]{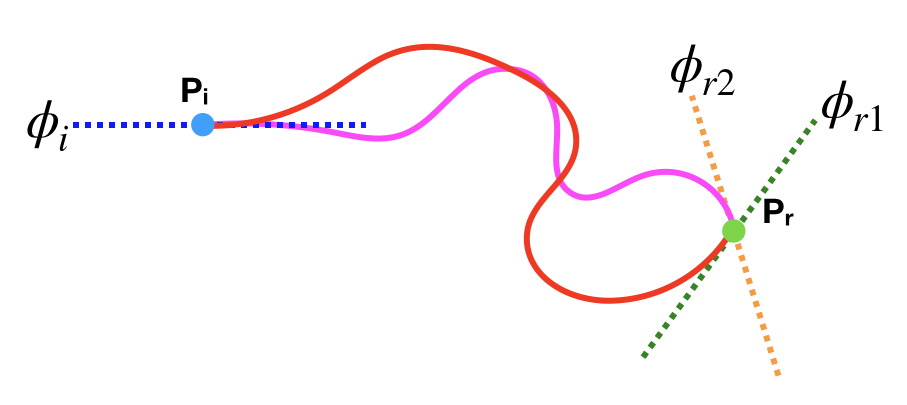}\\
    \caption{Schematic representation of the way spatial offset and the reconstruction of the polarization plane might be related. The red and purple curves represent two tracks originating from events having the same true impact point $\mathbf{P}_{\rm i}$, and being wrongly reconstructed at the track end in the same point $\mathbf{P}_{\rm r}$. $\phi_i$ is the polarization plane that would have been reconstructed as tangent to both tracks if the impacts points would have been correctly identified. $\phi_{r1}$ and $\phi_{r2}$ are the polarization planes reconstructed as tangent of the tracks in the point $\mathbf{P}_{\rm r}$.  }
    \label{fig:twotracks}
\end{figure}

The formalism we developed in Sect.~\ref{sec:linear} is based on the assumption that the displacement between the true interaction point $\mathbf{P}_{\rm i}$ and the reconstructed one $\mathbf{P}_{\rm r}$ is purely along the direction $\mathbf{n}$ of the reconstructed polarization plane (the photo-electron track). As a consequence, the determination of the reconstructed polarization plane is independent of the offset of the interaction point with respect to its true position (measuring correctly or not the interaction point does no change the reconstructed direction of the polarization plane). This is not in general true. The tracks can have a quite complex geometry, and in general the offset will have both a component along the reconstructed polarization plane (longitudinal), and one orthogonal to it (transverse). This implies not only that the offset is intrinsically 2D (as opposed to the 1D displacement assumption), but that even the reconstructed  polarization plane might be different if one locates correctly the impact point or not. Ideally the polarization plane should be the tangent plane to the track axis in the impact point. As can be seen from Fig.~\ref{fig:twotracks},  the tangent plane to a track axis can vary depending where the impact point is located. On top of this the rotation of the inferred polarization plane, might depend not just on the offset but also on track shape. This means that photons having all the same polarization properties, if the impact point is correctly reconstructed in $\mathbf{P}_{\rm i}$, will show different polarization properties when reconstructed in $\mathbf{P}_{\rm r}$. In this sense one cannot establish a one to one correspondence between the polarization at $\mathbf{P}_{\rm i}$ and that at $\mathbf{P}_{\rm r}$, and must resort to statistical arguments.\\
\\

In order to generalize the formalism, we assume that the chance an event reconstructed in $\mathbf{P}_{\rm r}$ originates in $\mathbf{P}_{\rm i}$, is higher if the reconstructed polarization plane in $\mathbf{P}_{\rm r}$, is aligned with the direction $\mathbf{P}_{\rm i} - \mathbf{P}_{\rm r}$ than if it is orthogonal. As a consequence if the distribution of the polarisation planes in the true impact points $\mathbf{P}_{\rm i}$ is uniform, then we expect that in the reconstructed point $\mathbf{P}_{\rm r}$, there should be an excess of photons with reconstructed polarization plane in the aligned direction:
\begin{align}
\mathcal{F}_{\rm r} (\phi) = \mathcal{F}_{\rm i} (\phi) [\mathcal{A}(\delta l) + \mathcal{B}(\delta l)\cos{(2(\phi-\phi_{\rm ir}))}]/2\pi
= \mathcal{F}_{\rm i} (\phi) [\mathcal{A}(\delta l)+ \mathcal{B}(\delta l)\cos{(2\phi)}\cos{(2\phi_{\rm ir})} +\mathcal{B}(\delta l)\sin{(2\phi)}\sin{(2\phi_{\rm ir})}]/2\pi 
\label{eq:dist}
\end{align}
where $\mathcal{F}_{\rm i} (\phi)$ is the distribution function of reconstructed polarization planes in the true impact point $\mathbf{P}_{\rm i}$, $\mathcal{A}(\delta l)$ represent the average (over polarization angle) chance an event is wrongly reconstructed at a distance $\delta l = {\rm dist}(\mathbf{P}_{\rm r},\mathbf{P}_{\rm i})$ from its true impact point, while  $\mathcal{B}(\delta l)$ (always $<\mathcal{A}(\delta l)$) accounts for the fact that the displacement chance is higher for event with a reconstructed polarization plane aligned with the direction $\mathbf{P}_{\rm i} - \mathbf{P}_{\rm r}$ corresponding to an angle $\phi_{\rm ir}$. $\mathcal{F}_{\rm r} (\phi)$ is then the distribution of events reconstructed in $\mathbf{P}_{\rm r}$ but originating in $\mathbf{P}_{\rm i}$. In Eq.~\ref{eq:dist}, we have assumed that the anisotropy in the distribution of the polarization planes can be described by a quadrupolar term. In general the full description might be more complex, however any higher harmonics, will have no effect of the determination of Stokes parameters, and for this reason we neglect it here.\\
\\
Let proceed as before assuming $\mathcal{F}_{\rm i} (\phi)$ is constant (unpolarized source). Then  $\mathcal{F}_{\rm i}$ is just the intensity at the point $\mathbf{P}_{\rm i}$. Assuming a point source of intensity $I_{\rm src}$ located at the origin of a reference frame, and a geometrical PSF $\mathcal{P}_{\rm g} (r)$ dependent only on distance from the origin, then one have:
\begin{align}
\mathcal{F}_{\rm i} = I_{\rm src}\mathcal{P}_{\rm g} (r_i)
\end{align}
where $r_i =\sqrt{x_i^2+y_i^2}$ and $\mathbf{P}_{\rm i} = (x_i,y_i)$. In terms of quantities measured at the reconstructed impact point $\mathbf{P}_{\rm r} = (x_r,y_r) = (x_i -\delta x, y_i -\delta y)$ one can approximate it as:
\begin{align}
\mathcal{F}_{\rm i} = I_{\rm src}\bigg[&\mathcal{P}_{\rm g} (r_r) + \frac{\partial \mathcal{P}_{\rm g}}{\partial x}\delta x + \frac{\partial \mathcal{P}_{\rm g}}{\partial y}\delta y +\frac{1}{2}\frac{\partial^2 \mathcal{P}_{\rm g}}{\partial x^2}\delta x^2 +\frac{1}{2}\frac{\partial^2 \mathcal{P}_{\rm g}}{\partial y^2}\delta y^2 
+\frac{\partial^2 \mathcal{P}_{\rm g}}{\partial x\partial y}\delta x\delta y  +\frac{1}{6}\frac{\partial^3 \mathcal{P}_{\rm g}}{\partial x^3}\delta x^3 +\frac{1}{6}\frac{\partial^3 \mathcal{P}_{\rm g}}{\partial y^3}\delta y^3 
+\frac{1}{2}\frac{\partial^2 \mathcal{P}_{\rm g}}{\partial x^2\partial y}\delta x^2\delta y +\nonumber\\
&+\frac{1}{2}\frac{\partial^2 \mathcal{P}_{\rm g}}{\partial x\partial y^2}\delta x\delta y^2 +\frac{1}{24}\frac{\partial^4 \mathcal{P}_{\rm g}}{\partial x^4}\delta x^4 +\frac{1}{24}\frac{\partial^4 \mathcal{P}_{\rm g}}{\partial y^4}\delta y^4 
+\frac{1}{6}\frac{\partial^4 \mathcal{P}_{\rm g}}{\partial x^3\partial y}\delta x^3\delta y +\frac{1}{6}\frac{\partial^4 \mathcal{P}_{\rm g}}{\partial x\partial y^3}\delta x\delta y^3 +\frac{1}{4}\frac{\partial^4 \mathcal{P}_{\rm g}}{\partial x\partial y^2}\delta x^2\delta y^2 \bigg] + \mathcal{O}(\delta l^3)
\end{align}
where derivatives are taken at $\mathbf{P}_{\rm r}$. With respect to the previous case, we now need to take derivatives up to the 4th order. On the other hand one has:
\begin{align}
\cos{(2\phi_{\rm ir})} &=  \cos^2{(\phi_{\rm ir})} -\sin^2{(\phi_{\rm ir})} = \frac{\delta x^2 - \delta y^2}{\delta l^2}\\    
   \sin{(2\phi_{\rm ir})} &=  2\cos{(\phi_{\rm ir})}\sin{(\phi_{\rm ir})} = 2\frac{\delta x\delta y}{\delta l^2}  
\end{align}
Hence:
\begin{align}
\mathcal{F}_{\rm r} (\phi) &= \frac{I_{\rm src}}{2\pi}\int_{-\infty}^{\infty} {\rm d}\delta x \int_{-\infty}^{\infty} {\rm d}\delta y\,\bigg\{ \mathcal{P}_{\rm g} \mathcal{A}(\delta l) +  \mathcal{A}(\delta l) \left(\dPgdx  \delta x +   \dPgdy \delta y\right)
 +\frac{1}{2}\mathcal{A}(\delta l) \left(\ddPgddx\delta x^2 +\ddPgddy\delta y^2\right)  +\nonumber\\
\quad & +\left(\mathcal{P}_{\rm g}\mathcal{B}(\delta l) \frac{\cos{(2\phi)}}{\delta l^2}   \right)(\delta x^2 -\delta y^2)+\left(\mathcal{A}(\delta l)\ddPgdxdy +2\mathcal{P}_{\rm g}\frac{\mathcal{B}(\delta l)}{\delta l^2}\sin{(2\phi)}  \right)\delta x\delta y +\nonumber\\
\quad &+\left(\frac{\mathcal{A}}{6}(\delta l)\dddPgdddx +\frac{\mathcal{B}(\delta l)}{\delta l^2}\dPgdx\cos{(2\phi)}  \right)\delta x^3+\left(\frac{\mathcal{A}(\delta l)}{6}\dddPgdddy -\frac{\mathcal{B}(\delta l)}{\delta l^2}\dPgdy  \right)\delta y^3+\nonumber\\
\quad &+\left(\frac{\mathcal{A}(\delta l)}{2}\dddPgddxdy + \frac{\mathcal{B}(\delta l)}{\delta l^2}\left[2\dPgdx\sin{(2\phi)} +\dPgdy \cos{(2\phi)} \right]  \right)\delta x^2\delta y+\nonumber\\
\quad &+\left(\frac{\mathcal{A}(\delta l)}{2}\dddPgdxddy + \frac{\mathcal{B}(\delta l)}{\delta l^2}\left[2\dPgdy\sin{(2\phi)} -\dPgdx \cos{(2\phi)} \right]   \right)\delta x\delta y^2+\nonumber\\
\quad &+
\left(\frac{\mathcal{A}(\delta l)}{24}\ddddPgddddx +\frac{\mathcal{B}(\delta l)}{2\delta l^2}\ddPgddx\cos{(2\phi)}   \right)\delta x^4
+
\left(\frac{\mathcal{A}(\delta l)}{24}\ddddPgddddy -\frac{\mathcal{B}(\delta l)}{2\delta l^2}\ddPgddy\cos{(2\phi)}\right)\delta y^4
+\nonumber\\
\quad &+
\left(\frac{\mathcal{A}(\delta l)}{6}\ddddPgdddxdy +\frac{\mathcal{B}(\delta l)}{\delta l^2}\left[\ddPgddx\sin{(2\phi)} +\ddPgdxdy\cos{(2\phi)} \right] \right)\delta x^3\delta y
+\nonumber\\
\quad &+
\left(\frac{\mathcal{A}(\delta l)}{6}\ddddPgdxdddy +\frac{\mathcal{B}(\delta l)}{\delta l^2}\left[\ddPgddy\sin{(2\phi)} -\ddPgdxdy\cos{(2\phi)} \right] \right)\delta x\delta y^3
+\nonumber\\
\quad &+\left(\frac{\mathcal{A}(\delta l)}{4}\ddddPgddxddy  +\frac{\mathcal{B}(\delta l)}{2\delta l^2}\left[4\ddPgdxdy\sin{(2\phi)}+\left(\ddPgddy-\ddPgddx\right)\cos{(2\phi)}
\right]
\right)\delta x^2\delta y^2
\end{align}
where the geometrical PSF $\mathcal{P}_{\rm g}$ and all its derivatives are evaluated in $\mathbf{P}_{\rm r}$. Note moreover that the anisotropy enters always through the term $\mathcal{B}(\delta l)/\delta l^2$. It is reasonable to expect that for  $\delta l=0$ one has $\mathcal{B}(\delta l)=0$, implying that when the impact point is correctly reconstructed, the distribution of the polarization planes of an unpolarized source has no preferential direction (but see \citep{Rankin_muler+22a} for a discussion on spurious modulation). Then symmetry considerations require that in the limit $\delta l\rightarrow 0$ one has $\mathcal{B}(\delta l)\propto \delta l^2$. We can then introduce a regularized distribution for the anisotropy $\tilde{\mathcal{B}}(\delta l)=\mathcal{B}(\delta l)/\delta l^2$ (note however that this regularization is ultimately not necessary, given that potentially diverging terms turn out to cancel each other). \\

Given that $\delta l =\sqrt{\delta x^2 + \delta y^2}$, the terms with odd powers in $\delta x$ and/or $\delta y$, cancel out, and only even terms remain. We can then define the following even moments related to the $\mathcal{A}(\delta l)$ distribution:
\begin{align}
\sigma_{\rm a}^2 &= \int_{-\infty}^\infty \int_{-\infty}^\infty \mathcal{A}(\delta l) \delta x^2 {\rm d} \delta x \,{\rm d} \delta y = \int_{-\infty}^\infty \int_{-\infty}^\infty \mathcal{A}(\delta l) \delta y^2 {\rm d} \delta x \,{\rm d} \delta y\\
\kappa_{\rm a} &= \int_{-\infty}^\infty \int_{-\infty}^\infty \mathcal{A}(\delta l) \delta x^4 {\rm d} \delta x \,{\rm d} \delta y = \int_{-\infty}^\infty \int_{-\infty}^\infty \mathcal{A}(\delta l) \delta y^4 {\rm d} \delta x \,{\rm d} \delta y\\
\eta_{\rm a} &= \int_{-\infty}^\infty \int_{-\infty}^\infty \mathcal{A}(\delta l) \delta x^2 \delta y^2 {\rm d} \delta x \,{\rm d} \delta y
\end{align}
and of the $\tilde{\mathcal{B}}(\delta l)$ distribution:
\begin{align}
\sigma_{\rm b}^2 &= \int_{-\infty}^\infty \int_{-\infty}^\infty \tilde{\mathcal{B}}(\delta l) \delta x^2 {\rm d} \delta x \,{\rm d} \delta y= \int_{-\infty}^\infty \int_{-\infty}^\infty \tilde{\mathcal{B}}(\delta l) \delta y^2 {\rm d} \delta x \,{\rm d} \delta y\\
\kappa_{\rm b} &= \int_{-\infty}^\infty \int_{-\infty}^\infty \tilde{\mathcal{B}}(\delta l) \delta x^4 {\rm d} \delta x \,{\rm d} \delta y= \int_{-\infty}^\infty \int_{-\infty}^\infty \tilde{\mathcal{B}}(\delta l) \delta y^4 {\rm d} \delta x \,{\rm d} \delta y\\
\eta_{\rm b} &= \int_{-\infty}^\infty \int_{-\infty}^\infty \tilde{\mathcal{B}}(\delta l) \delta x^2 \delta y^2 {\rm d} \delta x \,{\rm d} \delta y
\end{align}
\\
Now, using the following relation between cartesian and polar coordinates $\delta x = \delta l \cos{\theta}$, and  $\delta x = \delta l \sin{\theta}$ one  can show that:
\begin{align}
\int \int_{-\infty}^\infty \tilde{\mathcal{B}}(\delta l) \delta x^4 {\rm d} \delta x \,{\rm d} \delta y &= \int_{-\infty}^\infty \int_{-\infty}^\infty \tilde{\mathcal{B}}(\delta l) \delta l^4 \cos^4{(\theta)}\delta l\, {\rm d} \delta l \,{\rm d} \delta \theta   =  \frac{3\pi}{4}\int_{-\infty}^\infty \tilde{\mathcal{B}}(\delta l) \delta l^5 \, {\rm d} \delta l \\
\int \int_{-\infty}^\infty \tilde{\mathcal{B}}(\delta l) \delta x^2\delta y^2 {\rm d} \delta x \,{\rm d} \delta y &= \int_{-\infty}^\infty \int_{-\infty}^\infty \tilde{\mathcal{B}}(\delta l) \delta l^4 \cos^2{(\theta)}\sin^2{(\theta)}\delta l\, {\rm d} \delta l \,{\rm d} \delta \theta   =  \frac{\pi}{4}\int_{-\infty}^\infty \tilde{\mathcal{B}}(\delta l) \delta l^5 \, {\rm d} \delta l
\end{align}
which show that $\kappa = 3\eta$.
This leads to:
\begin{align}
\mathcal{F}_{\rm r} (\phi) &= \frac{I_{\rm src}}{2\pi}\bigg\{ \mathcal{P}_{\rm g} +\frac{\sigma_{\rm a}^2}{2} \left(\ddPgddx +\ddPgddy\right)  +\frac{\kappa_{\rm a}}{24}\left(\ddddPgddddx+\ddddPgddddy \right)+\frac{\eta_{\rm a}}{4} \ddddPgddxddy+\nonumber\\
\quad &+  \frac{\kappa_{\rm b}-\eta_{\rm b}}{2}\left(\ddPgddx-\ddPgddy \right)\cos{(2\phi)}
+2\eta_{\rm b}\ddPgdxdy\sin{(2\phi)}\bigg\}\\
&=\frac{I_{\rm src}}{2\pi}\bigg\{ \mathcal{P}_{\rm g} +\frac{\sigma_{\rm a}^2}{2} \left(\frac{\partial^2 \mathcal{P}_{\rm g}}{\partial r^2} +\frac{1}{r}\frac{\partial \mathcal{P}_{\rm g}}{\partial r}\right)  +\frac{\kappa_{\rm a}}{24}\left(\ddddPgddddx+\ddddPgddddy \right)+\frac{\kappa_{\rm a}}{12} \ddddPgddxddy+\nonumber\\
\quad &+  \left(\frac{\partial^2 \mathcal{P}_{\rm g}}{\partial r^2} -\frac{1}{r}\frac{\partial \mathcal{P}_{\rm g}}{\partial r}\right)\left(\frac{\kappa_{\rm b}-\eta_{\rm b}}{2}\frac{x^2-y^2}{r^2}\cos{(2\phi)}
+\eta_{\rm b}\frac{2xy}{r^2}\sin{(2\phi)}\right)\bigg\}\\
&=\frac{I_{\rm src}}{2\pi}\bigg\{ \mathcal{P}_{\rm g} +\frac{\sigma_{\rm a}^2}{2} \left(\frac{\partial^2 \mathcal{P}_{\rm g}}{\partial r^2} +\frac{1}{r}\frac{\partial \mathcal{P}_{\rm g}}{\partial r}\right)  +\frac{\kappa_{\rm a}}{24}\left(\ddddPgddddx+\ddddPgddddy +2 \ddddPgddxddy\right)+\nonumber\\
\quad &+  \left(\frac{\partial^2 \mathcal{P}_{\rm g}}{\partial r^2} -\frac{1}{r}\frac{\partial \mathcal{P}_{\rm g}}{\partial r}\right)\frac{\kappa_{\rm b}}{3}\left(\frac{x^2-y^2}{r^2}\cos{(2\phi)}
+\frac{2xy}{r^2}\sin{(2\phi)}\right)\bigg\}\\
&=\frac{I_{\rm src}}{2\pi}\bigg\{ \mathcal{P}_{\rm g} +\frac{\sigma_{\rm a}^2}{2} \left(\frac{\partial^2 \mathcal{P}_{\rm g}}{\partial r^2} +\frac{1}{r}\frac{\partial \mathcal{P}_{\rm g}}{\partial r}\right)  +\frac{\kappa_{\rm a}}{24}\left(\frac{\partial^4 \mathcal{P}_{\rm g}}{\partial r^4}+\frac{2}{r}\frac{\partial^3 \mathcal{P}_{\rm g}}{\partial r^3}-\frac{1}{r^2}\frac{\partial^2 \mathcal{P}_{\rm g}}{\partial r^2}-\frac{1}{r}\frac{\partial \mathcal{P}_{\rm g}}{\partial r}
\right)+\nonumber\\
\quad &+  \left(\frac{\partial^2 \mathcal{P}_{\rm g}}{\partial r^2} -\frac{1}{r}\frac{\partial \mathcal{P}_{\rm g}}{\partial r}\right)\frac{\kappa_{\rm b}}{3}\left(\frac{x^2-y^2}{r^2}\cos{(2\phi)}
+\frac{2xy}{r^2}\sin{(2\phi)}\right)\bigg\}
\end{align}
At this point we can proceed as before defining values of the Stokes parameters at the same point $\mathbf{P}_{\rm r} =(x,y)$ as:
\begin{align}
    I(x,y) = \int \mathcal{F}_{\rm r} (\phi) d\phi = I_{\rm src}\bigg\{ \mathcal{P}_{\rm g} +\frac{\sigma_{\rm a}^2}{2} \left(\frac{\partial^2 \mathcal{P}_{\rm g}}{\partial r^2} +\frac{1}{r}\frac{\partial \mathcal{P}_{\rm g}}{\partial r}\right)  +\frac{\kappa_{\rm a}}{24}\left(\frac{\partial^4 \mathcal{P}_{\rm g}}{\partial r^4}+\frac{2}{r}\frac{\partial^3 \mathcal{P}_{\rm g}}{\partial r^3}-\frac{1}{r^2}\frac{\partial^2 \mathcal{P}_{\rm g}}{\partial r^2}-\frac{1}{r}\frac{\partial \mathcal{P}_{\rm g}}{\partial r}
\right)\bigg\}
\end{align}
and:
\begin{align}
    Q(x,y) &= 2\int \mathcal{F}_{\rm r} (\phi)\cos{(2\phi)} d\phi = I_{\rm src}\frac{\kappa_{\rm b}}{3}\bigg\{
    \frac{\partial^2 \mathcal{P}_{\rm g}}{\partial r^2} -\frac{1}{r}\frac{\partial \mathcal{P}_{\rm g}}{\partial r}
    \bigg\}\frac{x^2-y^2}{r^2}\\
    U(x,y) &= 2\int \mathcal{F}_{\rm r} (\phi)\sin{(2\phi)} d\phi = I_{\rm src}\frac{\kappa_{\rm b}}{3}\bigg\{
    \frac{\partial^2 \mathcal{P}_{\rm g}}{\partial r^2} -\frac{1}{r}\frac{\partial \mathcal{P}_{\rm g}}{\partial r}
    \bigg\}\frac{2xy}{r^2}
\end{align}
One immediately sees that, for the effective PSF on intensity, apart from a higher order curvature term in $\kappa_{\rm a}$, the functional form is identical to our previous simplified derivation with the simple exchange $\sigma_{\rm f}^2 \rightarrow 2\sigma_{\rm a}^2$. For the PSF of the other Stokes parameter, that encode the radial profile of polarization induced by incorrect reconstruction of the track, again the functional form is identical to our previous simplified derivation with the simple exchange $\sigma_{\rm f}^2 \rightarrow 4\kappa_{\rm b}/3$. In particular if one assumes maximal anisotropy $\mathcal{B}(\delta l) = \mathcal{A}(\delta l)$, then, to first order in variance, one can show that $\kappa_{\rm b} = 3\sigma_{\rm a}^2/4$, and one recovers: \\
\begin{align}
    I(x,y) &= I_{\rm src}\bigg\{ \mathcal{P}_{\rm g} +\frac{\sigma_{\rm a}^2}{2} \left(\frac{\partial^2 \mathcal{P}_{\rm g}}{\partial r^2} +\frac{1}{r}\frac{\partial \mathcal{P}_{\rm g}}{\partial r}\right)\bigg\}\\
    Q(x,y) &=  I_{\rm src}\frac{\sigma_{\rm a}^2}{4}\bigg\{
    \frac{\partial^2 \mathcal{P}_{\rm g}}{\partial r^2} -\frac{1}{r}\frac{\partial \mathcal{P}_{\rm g}}{\partial r}
    \bigg\}\frac{x^2-y^2}{r^2}\\
    U(x,y) &=  I_{\rm src}\frac{\sigma_{\rm a}^2}{4}\bigg\{
    \frac{\partial^2 \mathcal{P}_{\rm g}}{\partial r^2} -\frac{1}{r}\frac{\partial \mathcal{P}_{\rm g}}{\partial r}
    \bigg\}\frac{2xy}{r^2}
\end{align}

\bibliographystyle{aa} 
\bibliography{MyBib.bib}

\end{document}